\documentclass[preprint,aps,prd,superscriptaddress,showpacs,nofootinbib]{revtex4}%
\usepackage{graphicx}
\usepackage{amsmath}
\usepackage{amssymb}
\usepackage{bm}
\usepackage{epsfig}
\newcommand{\intD}{\,\bm{\mathcal{N}}\hskip -2.945ex\int_{k}}
\newcommand{\intN}{\,\bm{\mathcal{N}}\hskip -2.95ex\int_{\bm{k}}} 
\newcommand{\intDtxt}{\,\bm{{\scriptscriptstyle%
\mathcal{N}}}\hskip -1.98ex\int_{\scriptscriptstyle k}}
\newcommand{\intNtxt}{\,\bm{{\scriptscriptstyle%
\mathcal{N}}}\hskip -1.98ex\int_{{\scriptscriptstyle \bm{k}}}}
\begin{document}
\preprint{ANL-HEP-PR-08-43}
\title{\mbox{}\\[10pt]
Order-$\bm{\alpha_s}$ corrections to the quarkonium electromagnetic current
\\
at all orders in the heavy-quark velocity}
\author{Geoffrey~T.~Bodwin}
\affiliation{High Energy Physics Division, Argonne National Laboratory,\\
9700 South Cass Avenue, Argonne, Illinois 60439, USA}
\author{Hee~Sok~Chung}
\affiliation{Department of Physics, Korea University, Seoul 136-701, Korea}
\author{Jungil~Lee}
\affiliation{High Energy Physics Division, Argonne National Laboratory,\\
9700 South Cass Avenue, Argonne, Illinois 60439, USA}
\affiliation{Department of Physics, Korea University, Seoul 136-701, Korea}
\author{Chaehyun~Yu}
\affiliation{Department of Physics, Korea University, Seoul 136-701, Korea}
\date{\today}
\begin{abstract}
We compute in order $\alpha_s$ the nonrelativistic QCD (NRQCD)
short-distance coefficients that match quark-antiquark operators of all
orders in the heavy-quark velocity $v$ to the electromagnetic current.
We employ a new method to compute the one-loop NRQCD contribution
to the matching condition. The new method uses full-QCD expressions
as a starting point to obtain the NRQCD contribution, thus greatly
streamlining the calculation. Our results show that, under a mild
constraint on the NRQCD operator matrix elements, the NRQCD velocity
expansion for the quark-antiquark-operator contributions to the
electromagnetic current converges. The velocity expansion converges
rapidly for approximate $J/\psi$ operator matrix elements.
\end{abstract}
\maketitle
\section{Introduction\label{sec:intro}}
The electromagnetic decays of quarkonia through a single virtual photon
have played an important r\^ole in the experimental and theoretical
development of quarkonium physics. On the experimental side, the decays
of charge-conjugation-odd quarkonium states to a lepton pair provide
unique signals for the detection of those states. On the theoretical
side, the decays of ${}^3S_1$ quarkonium states to a lepton pair 
allow one to determine one of the fundamental parameters of
the heavy-quark--antiquark ($Q\bar Q$) bound state, namely, the square
of the wave function at the origin. (See, for example,
Ref.~\cite{Bodwin:2007fz}.) The square of the wave function at the
origin enters into many calculations of quarkonium decay and production
rates.

The expression for the ${}^3S_1$ quarkonium decay rate into a lepton pair
at leading order in the QCD coupling $\alpha_s$ and at leading
order in $v$, the $Q$ or $\bar Q$ velocity in the quarkonium rest frame,
has been known since the first discovery of quarkonium and is based on
the Van Royen-Weisskopf formula \cite{Van Royen:1967nq} of quantum
electrodynamics. The order-$\alpha_s$ corrections to this formula at
leading order in $v$ were calculated 
in Refs.~\cite{Barbieri:1975ki,Celmaster:1978yz}. 
The order-$\alpha_s^0$ relativistic corrections
at relative orders $v^2$ and $v^4$ were calculated in
Refs.~\cite{Bodwin:1994jh,Bodwin:2002hg}, respectively.
Order-$\alpha_s^2$ corrections to the decay rate were calculated in
Refs.~\cite{Czarnecki:1997vz,Beneke:1997jm}.  The correction to
the electromagnetic current of a quarkonium at relative order 
$\alpha_s v^2$ was calculated in Ref.~\cite{Luke:1997ys}.

In this paper, we calculate relativistic corrections to the quarkonium
electromagnetic current at order $\alpha_s$. We carry out our
calculation in the context of nonrelativistic QCD (NRQCD)
\cite{Bodwin:1994jh}. We obtain closed-form expressions whose Taylor-series 
expansions in $v$ give the short-distance coefficients for the
NRQCD $Q\bar Q$ operators, of all orders in $v$, that match to the
electromagnetic current. We do not consider $Q\bar Q$ operators that
contain gauge fields. Therefore, our operators are not gauge
invariant, and we evaluate their matrix elements in the Coulomb gauge.
In the Coulomb gauge, $Q\bar Q$ operators involving gauge fields first
contribute at relative order $v^4$. Our results confirm the calculation
at relative order $\alpha_s v^2$ in Ref.~\cite{Luke:1997ys}. Since the
corrections at relative order $\alpha_s v^2$ are not very significant
at the current level of precision of calculations of $^{3}S_1$ quarkonium
electromagnetic decay rates, we do not expect the order-$\alpha_s$
corrections at still higher orders in $v$ to be important numerically.

We present our calculation primarily as a demonstration of a new method
for computing the one-loop NRQCD contribution that enters into the
matching of NRQCD to full QCD. The direct computation of one-loop NRQCD
expressions to all orders in $v$ would be a formidable task, in that it
would require knowledge of the NRQCD interactions and
electromagnetic-current operators, their Born-level short-distance
coefficients, and their Feynman rules to all orders in $v$. Instead of
following the direct NRQCD approach, we note that NRQCD through infinite
order in $v$ is equivalent to QCD, but with the interactions rearranged
in an expansion in powers of $v$. Therefore, we can obtain the one-loop
NRQCD contribution by starting from full-QCD expressions and expanding
integrands in powers of momenta divided by the heavy-quark mass $m$ {\it
before} we carry out the dimensional regularization. In dimensional
regularization, this method is related to the method of regions
\cite{Beneke:1997zp}. We explain this relationship in
Sec.~\ref{sec:matching}. The method of regions has been used previously
at leading order in $v$ to compute  NRQCD short-distance coefficients
from  full-QCD expressions. (See for example,
Ref.~\cite{Beneke:1997jm}.)

Our results show that, under a mild constraint on the NRQCD operator
matrix elements, the NRQCD velocity expansion for the
quark-antiquark-operator contributions to the electromagnetic current
converges. The velocity expansion converges rapidly for
approximate $J/\psi$ operator matrix elements.

The remainder of this paper is organized as follows. In
Sec.~\ref{sec:matching} we discuss the one-loop matching of NRQCD to QCD
at all orders in $v$. We define the notation that we use to describe the
kinematics of the calculation in Sec.~\ref{sec:kinematics}.
Section~\ref{sec:FORMCoeff} contains detailed formulas for the NRQCD $Q\bar Q$
short-distance coefficients. In Sec.~\ref{sec:QCD} we compute the
one-loop QCD corrections to the electromagnetic current, while in
Sec.~\ref{sec:NRQCD} we use our new method to compute the one-loop NRQCD
corrections to the electromagnetic current. We give analytic and
numerical results for the short-distance coefficients in
Sec.~\ref{sec:RESCoeff}, present a formula that resums a class
of relativistic corrections to all orders in $v$, and discuss the
convergence of the velocity expansion. Our conclusions are
given in Sec.~\ref{sec:Conclusion}. The Appendices contain compilations of
integrals and identities that are useful in the calculation.
\section{Matching to all orders in $\bm{v}$\label{sec:matching}}
We define the hadronic part of the quarkonium electromagnetic decay 
amplitude ${\cal A}_H^\mu$ as
\begin{equation}
(-iee_Q)i{\cal A}_H^\mu=\langle 0|J^\mu_{\rm EM}|H\rangle,
\end{equation}
where $H$ is the quarkonium, $e$ is the electromagnetic charge, $e_Q$ is 
the heavy-quark charge, and $J^\mu_{\rm EM}$ is the heavy-quark 
electromagnetic current:
\begin{equation}
J^\mu_{\rm EM}=(-iee_Q)\bar\psi\gamma^\mu\psi.
\end{equation}
Here, $\psi$ is the heavy-quark Dirac field, and $\gamma^\mu$ is a Dirac 
matrix. 

In the quarkonium rest frame,  $i{\cal A}_H^0=0$ because of conservation of 
the electromagnetic current. According to NRQCD 
factorization~\cite{Bodwin:1994jh}, we can write the spatial components 
$i{\cal A}_H^i$ as
\begin{equation}
\label{NRQCD-fact}%
i{\cal A}_H^i=\sqrt{2m_H}\sum_n c_n\langle 0|{\cal O}^i_n|H\rangle,
\end{equation}
where the $c_n$ are short-distance coefficients, the ${\cal O}^i_n$ are
NRQCD operators, and $m_H$ is the quarkonium mass. We regulate
the operator matrix elements in Eq.~(\ref{NRQCD-fact}) 
dimensionally in $d=4-2\epsilon$ dimensions. The
factor $\sqrt{2m_H}$ on the right side of Eq.~(\ref{NRQCD-fact})
appears because the NRQCD operator matrix elements have nonrelativistic
normalization, while we choose the amplitude on the left side of
Eq.~(\ref{NRQCD-fact}) to have relativistic normalization for the
quarkonium $H$.

The aim of this paper is to calculate the short-distance coefficients 
$c_n$ that correspond to $Q\bar Q$ color-singlet operators 
in order $\alpha_s^1$. We can determine these $c_n$ by making use of 
a matching equation that is the statement of NRQCD factorization for 
perturbative $Q\bar Q$ color-singlet states:
\begin{equation}
i{\cal A}_{Q\bar Q_1}^i=\sum_n c_n\langle 0|{\cal O}^i_n|Q\bar 
Q_1\rangle,
\end{equation}
where the subscript $1$ indicates a color-singlet state. 
Throughout this paper, we suppress the factor $\sqrt{N_c}$ 
that comes from the implicit color trace in $i{\cal A}_{Q\bar Q_1}^i$,
where $N_c = 3$ is the number of colors.
Through order $\alpha_s^1$, the matching equation is
\begin{equation}
\label{matching-01}%
i{\cal A}_{Q\bar Q_1}^{i(0)}+i{\cal A}_{Q\bar Q_1}^{i(1)}=
\sum_n (c_n^{(0)}+c_n^{(1)})\langle 0|{\cal O}^i_n|Q\bar Q_1\rangle^{(0)}
+\sum_n c_n^{(0)}\langle 0|{\cal O}^i_n|Q\bar Q_1\rangle^{(1)},
\end{equation}
where the superscripts $(0)$ and $(1)$ indicate the order in
$\alpha_s$. In the first sum in Eq.~(\ref{matching-01}), only
color-singlet $Q \bar Q$ operators contribute, while in the last sum,
additional operators can contribute if they mix into color-singlet
$Q\bar Q$ operators under one-loop corrections.

We define the quantity
\begin{equation}
\left[i{\cal A}_{Q\bar Q_1}^{i(0)}\right]_{\rm NRQCD} =
\sum_n c_n^{(0)}\langle 0|{\cal O}^i_n|Q\bar Q_1\rangle^{(0)},
\end{equation}
which is the expansion of $i{\cal A}_{Q\bar Q_1}^{i(0)}$ in powers of $q/m$, 
where $q$ is half the relative momentum of the heavy quark and heavy antiquark.
At order $\alpha_s^0$, the matching equation (\ref{matching-01}) yields
\begin{equation}
\label{match-0}%
i{\cal A}_{Q\bar Q_1}^{i(0)}=
\sum_n c_n^{(0)}\langle 0|{\cal O}^i_n|Q\bar Q_1\rangle^{(0)},
\end{equation}
from which the $c_n^{(0)}$ can be determined. The $c_n^{(0)}$ have 
been computed previously in Ref.~\cite{Bodwin:2007fz}.

At order $\alpha_s^1$, the matching equation (\ref{matching-01}) yields
\begin{equation}
\label{match-1}%
i{\cal A}_{Q\bar Q_1}^{i(1)}=
\sum_n c_n^{(1)}\langle 0|{\cal O}^i_n|Q\bar Q_1\rangle^{(0)}+
\left[i{\cal A}_{Q\bar Q_1}^{i(1)}\right]_{\rm NRQCD},
\end{equation}
from which the $c_n^{(1)}$ can be computed. 
We compute the quantities $i{\cal A}_{Q\bar Q_1}^{i(1)}$ and 
$\left[i{\cal A}_{Q\bar Q_1}^{i(1)}\right]_{\rm NRQCD}$
in Secs.~\ref{sec:QCD} and \ref{sec:NRQCD}, respectively. 

The quantity
\begin{equation}                                                      
\left[i{\cal A}_{Q\bar Q_1}^{i(1)}\right]_{\rm NRQCD} =       
\sum_n c_n^{(0)}\langle 0|{\cal O}^i_n|Q\bar Q_1\rangle^{(1)}          
\end{equation}
would be formidable to calculate directly in NRQCD because it involves
operators and interactions of all orders in $v$.
Rather than carry out such a direct calculation, we take a new
approach. We note that, by construction, NRQCD reproduces all of the
interactions in full QCD, but with those interactions reorganized in an
expansion in powers of $v$. Therefore, we can obtain $\left[i{\cal
A}_{Q\bar Q_1}^{i(1)}\right]_{\rm NRQCD}$ from the expression for $i{\cal
A}_{Q\bar Q_1}^{i(1)}$ by expanding the integrand in powers of the
momentum divided by $m$. 
Before making this expansion, we carry out the integration over the
temporal component of the loop momentum, using contour integration. This
procedure establishes the scale of the temporal component of the loop
momentum, which varies from contribution to contribution, and it avoids
the generation of ill-defined pinch singularities that can arise when
one expands the $Q$ and $\bar Q$ propagators prematurely in powers of 
the momentum.
We expand the integrand in powers of both the external momenta divided
by $m$ and the loop momenta divided by $m$. We then regulate the
integrals dimensionally, setting scaleless, power-divergent integrals
equal to zero. Ultimately, we renormalize ultraviolet divergences
according to the $\overline{\rm MS}$ prescription.

The procedure of expanding both external and loop momenta in power
series before regulating dimensionally was first utilized in the
Appendix of Ref.~\cite{Bodwin:1994jh}. The rationale for it was
discussed in Refs.~\cite{Luke:1997ys,Bodwin:1998mn}. This procedure
amounts to the prescription that infrared-finite contributions that
arise from loop momenta in the vicinity of zero are kept in the
short-distance coefficients \cite{Bodwin:1998mn}.\footnote{In the case
of hard-cutoff regularization, such as lattice regularization, the
expansion in powers of loop momentum divided by $m$ would be uniformly
convergent and would yield the same result as the unexpanded
expression.}

If one uses dimensional regularization, then the quantity $\sum_n
c_n^{(1)}\langle 0|{\cal O}^i_n|Q\bar Q_1\rangle^{(0)}$ corresponds to
the contribution from the hard region in the method of regions
\cite{Beneke:1997zp}, while the quantity $\left[i{\cal A}_{Q\bar
Q_1}^{i(1)}\right]_{\rm NRQCD}$ corresponds to the sum of the
contributions from the potential, soft, and ultrasoft regions, 
i.e., the contribution from the small-loop-momentum region. In the
method of regions, it is assumed that there are no contributions from
the region in which the temporal component of the gluon momentum is of
order $m$, but the spatial component of the gluon momentum is of order
$mv$. As we shall see explicitly in our calculation, this assumption is
justified because the contribution from this region of integration vanishes
in dimensional regularization. We note, however, that the contribution
from this region does not vanish in the case of a hard-cutoff regulator.
One potentially useful feature of the approach that we present here is
that it can be applied in the case of a hard cutoff, such as lattice
regularization, while the method of regions is applicable only in
dimensional regularization. In the method of regions, one can compute the
contribution from the hard region directly, rather than computing it, as
we do, by subtracting the small-loop-momentum contribution from the
full-QCD contribution. As we shall explain later, there may be
advantages to our indirect procedure in calculating the hard
contribution to all orders in $v$.

\section{Kinematics\label{sec:kinematics}} 
Before proceeding to write explicit formulas for the short-distance
coefficients, let us define some notation for the kinematics of the
heavy-quark electromagnetic vertex. We take $p_1$ and $p_2$ to be the momenta
of the incoming heavy quark $Q$ and heavy antiquark $\bar{Q}$, respectively.
$p_1$ and $p_2$ can be expressed as linear combinations of their
average $p$ and half their difference $q$:
\begin{subequations}
\label{momenta-definition}%
\begin{eqnarray}
p_1&=&p+q,
\\
p_2&=&p-q.
\end{eqnarray}
\end{subequations}
In the $Q\bar{Q}$ rest frame, the momenta are given by
\begin{subequations}
\label{momenta-rest}%
\begin{eqnarray}
p_1&=&(E,\bm{q}),
\\
p_2&=&(E, -\bm{q}),
\\
p&=&(E,\bm{0}),
\\
q&=&(0,\bm{q}),
\end{eqnarray}
\end{subequations}
where $E=\sqrt{m^2+\bm{q}^2}$.
The quark $Q$ and antiquark $\bar{Q}$ are on their mass shells: 
$p_1^2=p_2^2=m^2$. For later use,
it is convenient to define a parameter
\begin{equation}
\label{delta}%
\delta=\frac{|\bm{q}|}{E},
\end{equation}
which is related to the velocity
\begin{equation}
\label{v}%
v=\frac{|\bm{q}|}{m}.
\end{equation}
We can write $\delta$ in terms of $v$ as
\begin{equation}
\label{delta-v}%
\delta=\frac{v}{\sqrt{1+v^2}}.
\end{equation}
$E^2$ and $\bm{q}^2$ are expressed in terms of $m$ and $\delta$ as
\begin{subequations}
\label{eq-delta}%
\begin{eqnarray}
E^2&=&\frac{m^2}{1-\delta^2},
\\
\bm{q}^2&=&\frac{m^2\delta^2}{1-\delta^2}.
\end{eqnarray}
\end{subequations}
\section{Formulas for the short-distance coefficients\label{sec:FORMCoeff}}
Now let us make use of the matching conditions (\ref{match-0}) and 
(\ref{match-1}) to compute the short-distance coefficients for the 
specific color-singlet $Q\bar Q$ operators that we consider in this 
paper. These operators are
\begin{subequations}
\label{OPerators}%
\begin{eqnarray}
\label{OPeratorsA}%
{\cal O}_{An}^i&=&\chi^\dagger (-\tfrac{i}{2}\tensor{\bm{\nabla}})^{2n}
\sigma^i\psi,
\\
{\cal O}_{Bn}^i&=&\chi^\dagger (-\tfrac{i}{2}\tensor{\bm{\nabla}})^{2n-2}
(-\tfrac{i}{2}\tensor{\nabla}^i)
(-\tfrac{i}{2}\tensor{\bm{\nabla}})\cdot\bm{\sigma}
\psi,
\end{eqnarray}
\end{subequations}
where $\psi$ is the Pauli spinor field that annihilates a heavy quark,  
$\chi^\dagger$ is the Pauli spinor field that annihilates a heavy antiquark, 
and $\sigma^i$ is a Pauli matrix. 
Our operators contain ordinary derivatives, rather 
than covariant derivatives. Therefore, our operators are not gauge invariant, 
and we evaluate their matrix elements in the Coulomb gauge. 
We do not consider $Q\bar Q$ operators involving the gauge fields,
which first contribute at relative order $v^4$.
We note that ${\cal O}_{Bn}^i$ can be decomposed into a linear combination
of the $S$-wave operator ${\cal O}_{An}^i$ and the $D$-wave operator
${\cal O}_{Dn}^i$:
\begin{equation}
\label{OPerator-B}%
{\cal O}_{Bn}^i=
\frac{1}{d-1}
{\cal O}_{An}^i+
{\cal O}_{Dn}^i,
\end{equation}
where ${\cal O}_{Dn}^i$ is defined by 
\begin{equation}
\label{OPerator-D}%
{\cal O}_{Dn}^i=
\chi^\dagger (-\tfrac{i}{2}\tensor{\bm{\nabla}})^{2n-2}
\left[
(-\tfrac{i}{2}\tensor{\nabla}^i)
(-\tfrac{i}{2}\tensor{\bm{\nabla}})\cdot\bm{\sigma}
-\frac{1}{d-1}
(-\tfrac{i}{2}\tensor{\bm{\nabla}})^{2}
\sigma^i
\right]
\psi.
\end{equation}
In the basis of operators ${\cal O}^i_{An}$ and ${\cal O}^i_{Bn}$,
the matching conditions (\ref{match-0}) and (\ref{match-1}) become
\begin{subequations}
\label{coeffs-a-b}%
\begin{eqnarray}
\label{coeffs-a-b-0}%
i{\cal A}_{Q\bar Q_1}^{i(0)}
&=&
\sum_n a_n^{(0)}\langle 0|{\cal O}^i_{An}|Q\bar Q_1\rangle^{(0)}
+
\sum_n b_n^{(0)}\langle 0|{\cal O}^i_{Bn}|Q\bar Q_1\rangle^{(0)},
\\
\label{coeffs-a-b-1}%
i{\cal A}_{Q\bar Q_1}^{i(1)}
&=&
\sum_n a_n^{(1)}\langle 0|{\cal O}^i_{An}|Q\bar Q_1\rangle^{(0)}
+
\sum_n b_n^{(1)}\langle 0|{\cal O}^i_{Bn}|Q\bar Q_1\rangle^{(0)}
+
\left[i{\cal A}_{Q\bar Q_1}^{i(1)}\right]_{\rm NRQCD},
\end{eqnarray}
\end{subequations}
where $a_n$ and $b_n$ are the corresponding short-distance coefficients.
A similar equation holds in the basis 
${\cal O}^i_{An}$ and ${\cal O}^i_{Dn}$, where the associated 
short-distance coefficients are
\begin{subequations}
\label{coeffs-sd}%
\begin{eqnarray}
\label{coeffs-s}
s_n&=&a_n+\frac{1}{d-1}\,b_n,
\\
d_n&=&b_n,
\end{eqnarray}
\end{subequations}
respectively.\footnote{
If we replace the ordinary derivatives $\tensor{\bm{\nabla}}$ with
covariant derivatives $\tensor{\bm{D}}$ in an $S$-wave operator ${\cal
O}_{An}^i$, then we obtain one of the conventional gauge-invariant
$S$-wave NRQCD operators. Because the squared covariant derivatives
$(\tensor{\bm{D}})^2$ commute with themselves, the substitution of
covariant derivatives for ordinary derivatives leads to a unique $S$-wave
operator at each order $n$. Therefore, the $S$-wave short-distance
coefficients $s_n$ that we compute are also the short-distance
coefficients of the $S$-wave operator in which ordinary derivatives have
been replaced with covariant derivatives. In the case of the $D$-wave
operators ${\cal O}_{Dn}^i$, the replacement of ordinary derivatives
with covariant derivatives does not lead to a unique operator because
$(\tensor{D})^i$ and $(\tensor{D})^j$ do not commute. Therefore, each of the
$D$-wave short-distance coefficients $d_n$ that we compute is the sum of
the short-distance coefficients for the various operators at order $n$
that can be constructed from covariant derivatives.}

The $Q\bar Q$ matrix elements in Eq.~(\ref{coeffs-a-b}) are
\begin{subequations}
\label{pert-me}%
\begin{eqnarray}
\langle 0|{\cal O}_{An}^i|Q\bar Q_1\rangle^{(0)}
&=&\bm{q}^{2n}\eta^\dagger \sigma^i \xi ,\\
\langle 0|{\cal O}_{Bn}^i|Q\bar Q_1\rangle^{(0)}
&=&\bm{q}^{2n-2}q^i\eta^\dagger\bm{q}\cdot\bm{\sigma}\xi,
\end{eqnarray}
\end{subequations}
where $\xi$ and $\eta$ are two-component spinors. 
In order to maintain consistency with our calculations in full 
QCD, we have taken the $Q\bar Q$ states to have nonrelativistic 
normalization and we have suppressed the factor $\sqrt{N_c}$
that comes from the color trace. 

Because of current conservation, the most general form of 
$i \mathcal{A}_{Q\bar{Q}_1}^i$ is 
\begin{equation}
i \mathcal{A}_{Q\bar{Q}_1}^i = 
\bar{v} (p_2) (G \gamma^i + H q^i) u(p_1),
\end{equation}
where 
\begin{equation}
\label{A-Z-L}%
G =Z_Q (1+\Lambda).
\end{equation}
$Z_Q$ is the fermion wave-function renormalization, and
$\Lambda$ is the multiplicative correction to the fermion
electromagnetic vertex. Similarly, 
\begin{equation}
i \left[\mathcal{A}_{Q\bar{Q}_1}^i\right]_{\textrm{NRQCD}} =
\bar{v} (p_2) (G_{\textrm{NRQCD}} \gamma^i 
             + H_{\textrm{NRQCD}} q^i) u(p_1),
\end{equation}
where
\begin{equation}
\label{GNRQCD}
G_{\textrm{NRQCD}}=[Z_Q]_\textrm{NRQCD} (1+\Lambda_\textrm{NRQCD}).
\end{equation}

Using nonrelativistic normalization for the spinors $u$ and $v$, we obtain
\begin{subequations}
\label{spinor-reduction}%
\begin{eqnarray}
\bar{v}(p_2) \gamma^i u(p_1)
&=& \eta^\dagger \sigma^i \xi
- \frac{q^i \eta^\dagger \bm{q}\cdot\bm{\sigma} \xi}{E (E+m)},
\\
q^i \bar{v}(p_2) u(p_1) &=&
-\,\, \frac{q^i\eta^\dagger \bm{q}\cdot\bm{\sigma} \xi}{E}.
\end{eqnarray}
\end{subequations}
Then,
\begin{equation}
\label{amp-A-B}%
i{\cal A }^i_{Q\bar Q_1}= G\eta^\dagger\sigma^i\xi
-\left[\frac{ G}{E(E+m)}+\frac{ H }{E}\,\right]
q^i\eta^\dagger\bm{q}\cdot\bm{\sigma}\xi.
\end{equation}
Similarly,
\begin{equation}
\label{amp-A-B-NRQCD}%
i\left[{\cal A }^i_{Q\bar Q_{1}}\right]_{\textrm{NRQCD}}
= G_{\textrm{NRQCD}}\eta^\dagger\sigma^i\xi
-\left[\frac{ G_{\textrm{NRQCD}}}{E(E+m)}
      +\frac{ H_{\textrm{NRQCD}}}{E}\,\right]
q^i\eta^\dagger\bm{q}\cdot\bm{\sigma}\xi.
\end{equation}

Using the matching condition (\ref{coeffs-a-b-0}) and 
Eqs.~(\ref{pert-me}) and (\ref{amp-A-B}), we obtain
the short-distance coefficients at order $\alpha_s^0$:
\begin{subequations}
\label{an0-bn0}%
\begin{eqnarray}
a_n^{(0)}&=&
\left.
\frac{1}{n!}\left(\frac{\partial}{\partial \bm{q}^2}\right)^n
G^{(0)}
\right|_{\bm{q}^2=0}
=
\delta_{n0},\\
b_n^{(0)}&=&d^{(0)}_n=
-
\left.
\frac{1}{(n-1)!}\left(\frac{\partial}{\partial \bm{q}^2}\right)^{n-1}
\left[\frac{ G^{(0)}}{E(E+m)}
+\frac{ H^{(0)}}{E}\right]
\right|_{\bm{q}^2=0}
\nonumber\\
&=&
-
\left.
\frac{1}{(n-1)!}\left(\frac{\partial}{\partial 
\bm{q}^2}\right)^{n-1}\left[\frac{1}{E(E+m)}\right]
\right|_{\bm{q}^2=0},
\\
s_n^{(0)}&=& a_n^{(0)}+\frac{1}{3}\,b_n^{(0)}.
\end{eqnarray}
\end{subequations}
Using the matching condition (\ref{coeffs-a-b-1}) and
Eqs.~(\ref{pert-me}), (\ref{amp-A-B}), and (\ref{amp-A-B-NRQCD}),
we obtain the short-distance coefficients at
order $\alpha_s^1$:
\begin{subequations}
\begin{eqnarray}
a_n^{(1)}&=&
\left.
\frac{1}{n!}
\left(\frac{\partial}{\partial \bm{q}^2}\right)^n
\Delta G^{(1)}
\right|_{\bm{q}^2=0}
,\\
b_n^{(1)}&=&  d_n^{(1)}= 
-
\left.
\frac{1}{(n-1)!}\left(\frac{\partial}{\partial \bm{q}^2}\right)^{n-1}
\left[\frac{\Delta G^{(1)}}{E(E+m)}
+\frac{\Delta H^{(1)}}{E}
\right]
\right|_{\bm{q}^2=0}
,
\\
\label{dm1}
s_n^{(1)}&=& a_n^{(1)}+\frac{1}{d-1}\,b_n^{(1)},
\end{eqnarray}
\end{subequations}
where
\begin{subequations}
\label{DeltaAB}%
\begin{eqnarray}
\Delta G^{(1)}&=&
G^{(1)}- G_{\rm NRQCD}^{(1)},\\
\Delta H^{(1)}&=&
H^{(1)}- H_{\rm NRQCD}^{(1)}.
\end{eqnarray}
\end{subequations}

The infrared divergences in
 $G^{(1)}_{\textrm{NRQCD}}$ and $H^{(1)}_{\textrm{NRQCD}}$
cancel in 
$\Delta G^{(1)}$ and $\Delta H^{(1)}$
because NRQCD reproduces full QCD in the infrared region.
The one-loop NRQCD matrix elements in $G_{\rm NRQCD}^{(1)}$ contain 
ultraviolet divergences, which we renormalize according to the 
$\overline{\rm MS}$ prescription. The quantity $H_{\rm NRQCD}^{(1)}$ 
is free of ultraviolet divergences. 
The quantities $\Lambda$ and $Z_Q$ 
also contain ultraviolet divergences. However, because of the usual 
cancellation between the vertex and fermion-wave-function
renormalizations, $G^{(1)}$ is free of ultraviolet divergences. 
$H^{(1)}$ is also free of ultraviolet divergences. Carrying out the 
renormalization, we have
\begin{subequations}
\label{ab-MSbar}%
\begin{eqnarray}
\left[a_n^{(1)}\right]_{\overline{\rm MS}}&=&
\left.
\frac{1}{n!}
\left(\frac{\partial}{\partial \bm{q}^2}\right)^n
\Delta G^{(1)}_{\overline{\rm MS}}
\,\right|_{\bm{q}^2=0}
,
\\
\left[b_n^{(1)}\right]_{\overline{\rm MS}}&=&
\left[d_n^{(1)}\right]_{\overline{\rm MS}}=
-
\left.
\frac{1}{(n-1)!}\left(\frac{\partial}{\partial \bm{q}^2}\right)^{n-1}
\left[\frac{\Delta G^{(1)}_{\overline{\rm MS}}}{E(E+m)}
+\frac{\Delta H^{(1)}}{E}
\right]
\right|_{\bm{q}^2=0},
\\
\label{dm1MSbar}
\left[s_n^{(1)}\right]_{\overline{\rm MS}}&=&
\left[a_n^{(1)}\right]_{\overline{\rm MS}}+
\frac{1}{3}\left[b_n^{(1)}\right]_{\overline{\rm MS}}.
\end{eqnarray}
\end{subequations}
In deriving the expression for $\left[s_n^{(1)}\right]_{\overline{\rm MS}}$,
we have used the fact that,
in minimal subtraction, one removes the $1/\epsilon$ pole times the
order-$\alpha_s^0$ $d$-dimensional matrix element. Hence, a term proportional
to $(d-1)^{-1}\epsilon^{-1}$ is subtracted in Eq.~(\ref{dm1}) in carrying out
the renormalization.
\section{QCD Corrections\label{sec:QCD}}
In this section, we calculate the QCD corrections to the heavy-quark 
electromagnetic current.
That is, we compute $i{\cal A}_{Q\bar Q_1}^{i(1)}$.
\subsection{Vertex Correction \label{subsec:QCDvertex}} 
In the Feynman gauge, the vertex correction to the electromagnetic current
is given by
\begin{equation}
\label{v-amp}%
\Lambda^\mu
=
-ig_s^2C_F
\int_k
\frac{
\bar{v}(p_2)
\gamma_\alpha
(-/\!\!\!p_2+/\!\!\!k+m)
\gamma^\mu
(/\!\!\!p_1+/\!\!\!k+m)
\gamma^\alpha
u(p_1)
}
{D_0D_1D_2},
\end{equation}
where $g_s^2=4\pi\alpha_s$ is the strong coupling,
$C_F=(N_c^2-1)/(2N_c)=4/3$, and
\begin{subequations}
\label{Di-definition}%
\begin{eqnarray}
\label{int-definition}%
\int_k
&\equiv&
\mu^{2\epsilon}
\int\frac{d^dk}{(2\pi)^d},
\\
D_0&=&k^2+i \varepsilon,
\\
D_1&=&k^2+2k\cdot p_1+i \varepsilon,
\\
D_2&=&k^2-2k\cdot p_2+i \varepsilon.
\end{eqnarray}
\end{subequations}
$\mu$ is the renormalization scale.
The loop momentum $k$ is chosen to be the gluon momentum.

By making use of Eq.~(\ref{momenta-definition}) and applying the equations
of motion, 
\begin{subequations}
\label{diraceq}%
\begin{eqnarray}
\bar{v}(p_2) /\!\!\!p u(p_1) &=& 0, \\
\bar{v}(p_2) /\!\!\!q u(p_1) &=& m \bar{v}(p_2) u(p_1),
\end{eqnarray}
\end{subequations}
we find that Eq.~(\ref{v-amp}) can be written as
\begin{eqnarray}
\label{numerator}%
\Lambda^\mu
&=&
-ig_s^2C_F
\int_k
\frac{1} {D_0D_1D_2}
\,
\bar{v} (p_2) \bigg\{ 
                      \bigg[(d-2)k^2 - 4(2 p^2 - m^2) + 8  k \cdot q  
                      \bigg]
                      \gamma^\mu 
\nonumber\\
&&\hspace{27ex}
+ 4 m k^{\mu} - 8 q^{\mu} k\!\!\!/ + 2 (2-d) k^{\mu} k\!\!\!/ 
\,\bigg\} u(p_1).
\end{eqnarray}
Tensor reductions of the integrals in
Eq.~(\ref{numerator}) are given in Appendix~\ref{appendix:tensor}.
The result is
\begin{eqnarray}
\label{lambdainS}%
\Lambda^{\mu} &=& -i g_s^2 C_F \bar{v}(p_2)\Bigg\{
\bigg[
(d-2) J_1 
-4 (2 p^2 - m^2) J_2 + 4 J_3 + 2 (2-d) J_4
\bigg] 
\gamma^{\mu} 
\nonumber \\
&&\hspace{15ex}
+ \frac{2 m p^{\mu}}{p^2} J_5 - \frac{2 m q^{\mu}}{q^2} J_3
+ 2 (2-d) m \left( \frac{q^{\mu}}{q^2} J_6 
+ \frac{p^{\mu}}{p^2 q^2} J_7 \right) \Bigg\}u(p_1),
\end{eqnarray}
where the integrals $J_i$ are defined by
\begin{equation}
\label{Si}%
J_i = \int_k \frac{N_i}{D_0 D_1 D_2},
\end{equation}
and
\begin{subequations}
\label{Ni-definition}%
\begin{eqnarray}
N_1 &=& k^2, \\ 
N_2 &=& 1, \\
N_3 &=& 2 k \cdot q, \\
N_4 &=&  
\frac{1}{d-2} \left[
k^2 
- \frac{(k \cdot p)^2}{p^2} 
- \frac{(k \cdot q)^2}{q^2} 
\right],\\
N_5 &=& 2 k \cdot p,\\ 
N_6 &=& \frac{1}{d-2} \left[
-k^2 
+ \frac{(k \cdot p)^2}{p^2} 
+ (d-1) \frac{(k \cdot q)^2}{q^2} 
\right]
, \\
N_7 &=& 
k \cdot p\,
k \cdot q.
\end{eqnarray}
\end{subequations}
The integrals $J_1$--$J_7$ are evaluated in 
Appendix~\ref{appendix:scalarintegrals}.
The results are tabulated in Eq.~(\ref{Ji-final}).
We note that $J_5$ and $J_7$ vanish, as is required by conservation of
electromagnetic current in Eq.~(\ref{lambdainS}).

Writing the vertex correction as 
$\Lambda^{\mu} =  \bar{v}(p_2)(\Lambda\gamma^{\mu} + H q^{\mu})u(p_1)$,
we have
\begin{subequations}
\label{Lam-final}%
\begin{eqnarray}
\label{Lam-finala}%
\Lambda &=& - i g_s^2 C_F \bigg[(d-2) J_1 - 4 (2 p^2 -m^2) J_2
+ 4 J_3 + 2 (2-d)J_4 \bigg] \nonumber \\
&=& \frac{\alpha_s C_F}{4 \pi}
\bigg\{
 \frac{1}{\epsilon_{\textrm{UV}}}
+\log \frac{4 \pi \mu^2 e^{-\gamma_{_{\!\textrm{E}}}}}{m^2}
+2(1+\delta^2) L(\delta) \left(
 \frac{1}{\epsilon_{\textrm{IR}}}
+\log \frac{4 \pi \mu^2 e^{-\gamma_{_{\!\textrm{E}}}}}{m^2}
 \right)
+ 6 \delta^2 L(\delta)
\nonumber\\
&&
\hspace{7ex}
- 4(1+\delta^2)K(\delta)
+(1+\delta^2)
\left[
\frac{\pi^2}{\delta}
- \frac{i \pi}{\delta}
\left(
\frac{1}{\epsilon_{\textrm{IR}}}
+
\log \frac{\pi \mu^2 e^{-\gamma_{_{\!\textrm{E}}}}}{\bm{q}^2}
+\frac{3 \delta^2}{1+\delta^2}
\right)
\right]
\bigg\},
\nonumber\\
\\
\label{Lam-finalb}%
H&=& -i g_s^2 C_F \left[ - \frac{2 m}{q^2} J_3
+ \frac{2 (2-d) m}{q^2} J_6 \right]=
\frac{\alpha_s C_F}{4 \pi} \frac{1-\delta^2}{m}
\left[ 2 L(\delta) - \frac{i \pi}{\delta} \right],
\end{eqnarray}
\end{subequations}
where the subscripts on $1/\epsilon$ denote the origins of the divergences
and $\gamma_{_{\!\textrm{E}}}$ is the Euler-Mascheroni constant.
The functions $L (\delta)$ and $K(\delta)$ are given by
\begin{subequations}
\label{L-K-delta}%
\begin{eqnarray}
\label{L-delta}%
L(\delta)&=& \frac{1}{2 \delta} 
\log \left( \frac{1 + \delta}{1 - \delta} \right), 
\\
\label{K-delta}%
K(\delta) &=& 
\frac{1}{4 \delta}
\left[ \textrm{Sp} \left( \frac{2 \delta}{1+\delta} \right)
- \textrm{Sp} \left( - \frac{2 \delta}{1-\delta} \right) 
\right],
\end{eqnarray}
\end{subequations}
where $\textrm{Sp}$ is the Spence function:
\begin{equation}
\label{Sp}%
\textrm{Sp}(x)=
\int_x^0\frac{\log (1-t)}{t}dt.
\end{equation}
In Eq.~(\ref{Lam-final}),  we have neglected terms of order $\epsilon^1$
and higher. In the remainder of this paper, we drop such higher-order terms. 
\subsection{Wave-function Renormalization\label{subsec:QCDZQ}}
The heavy-quark wave-function renormalization $Z_Q$,
evaluated in dimensional
regularization, is given in Ref.~\cite{Braaten:1995ej}:
\begin{equation}
\label{zq}%
Z_Q=
1+\frac{\alpha_s C_F}{4\pi}
\left(
-\frac{1}{\epsilon_{\textrm{UV}}}
-\frac{2}{\epsilon_{\textrm{IR}}}
-3
\log 
\frac{4 \pi \mu^2 e^{-\gamma_{_{\!\textrm{E}}}}}{m^2}
-4
\right).
\end{equation}
\subsection{ Summary of QCD results\label{subsec:QCDsummary}}
By making use of Eqs.~(\ref{A-Z-L}), (\ref{Lam-final}), and (\ref{zq}), 
we find that $G$ and $H$ are given by
\begin{subequations}
\label{QCDAB}%
\begin{eqnarray}
G
&=& 1 + \frac{\alpha_s C_F}{4 \pi} 
\bigg\{
2 \big[ (1+\delta^2) L (\delta) -1\big]
\left(
\frac{1}{\epsilon_{\textrm{IR}}}
+ 
\log \frac{4 \pi \mu^2 e^{-\gamma_{_{\!\textrm{E}}}}}{m^2}
\right)
+ 6 \delta^2 L(\delta) - 
4 (1+\delta^2) K(\delta) 
\nonumber \\
&&\hspace{11ex}
-4 
+
(1+\delta^2)
\left[ \frac{\pi^2}{\delta} - \frac{i \pi}{\delta} 
\left(
\frac{1}{\epsilon_{\textrm{IR}}} 
+ 
\log \frac{\pi \mu^2 e^{-\gamma_{_{\!\textrm{E}}}}}{\bm{q}^2}
+\frac{3 \delta^2}{1+\delta^2} 
\right)
\right]
\bigg\},
\\
H &=& \frac{\alpha_s C_F}{4 \pi} \frac{1-\delta^2}{m} 
\left[ 2 L(\delta) - \frac{i \pi}{\delta} \right].
\end{eqnarray}
\end{subequations}

Expanding Eq.~(\ref{amp-A-B}) through order $v^2$, using Eq.~(\ref{QCDAB}),
we obtain
\begin{eqnarray}
\label{amp-qq-v2}%
i \mathcal{A}^i_{Q\bar{Q}_{1}}
&=& \eta^\dagger \sigma^i \xi
\Bigg[  1 + \frac{\alpha_s C_F}{4 \pi}
\bigg\{ \frac{8}{3} v^2 
       \left(  
             \frac{1}{\epsilon_{\textrm{IR}}} 
      + \log \frac{4 \pi \mu^2 e^{-\gamma_{_{\!\textrm{E}}}}}{m^2} 
       \right) 
-8 + \frac{2v^2}{9}
\nonumber \\
&& \hspace{17ex} 
+ \left( 1 + \frac{3v^2}{2} \right) 
\left[
       \frac{\pi^2}{v} 
     - \frac{i \pi}{v} \left(
                       \frac{1}{\epsilon_{\textrm{IR}}} 
                       + \log \frac{\pi \mu^2 e^{-\gamma_{_{\!\textrm{E}}}}}
                                   {\bm{q}^2} 
                       \right) 
\right] 
-3 i \pi v
\bigg\}
\Bigg]
\nonumber \\
&&
- \frac{q^i \eta^\dagger \bm{q}\cdot\bm{\sigma} \xi}{2 m^2}
\Bigg\{ 1 + \frac{\alpha_s C_F}{4 \pi}
\bigg[ -4 + \frac{\pi^2}{v} 
     - \frac{i \pi}{v}\left(
     \frac{1}{\epsilon_{\textrm{IR}}} 
       + \log \frac{\pi \mu^2 e^{-\gamma_{_{\!\textrm{E}}}}}
                   {\bm{q}^2} 
       +2
                      \right)
\bigg] \Bigg\}
\nonumber\\
&&+O(v^{3}).
\end{eqnarray}
Equation~(\ref{amp-qq-v2}) agrees with Eq.~(4.16) of Ref.~\cite{Luke:1997ys}.
\section{NRQCD corrections\label{sec:NRQCD}}
In this section, we calculate the NRQCD corrections to the heavy-quark
electromagnetic current.
That is, we compute 
$\left[i{\cal A}_{Q\bar Q_1}^{i(1)}\right]_{\rm NRQCD}$.
In order to demonstrate our method for calculating
these corrections from full-QCD expressions, we present the calculation
in some detail.

Divergent integrals are regulated using dimensional regularization,
with $d=4-2\epsilon$.
We define the following notation for the loop integrals in $d-1$ dimensions:
\begin{equation}
\label{int3-definition}%
\int_{\bm{k}}\equiv
\mu^{2 \epsilon} \int \frac{d^{d-1}k}{(2 \pi)^{d-1}}.
\end{equation}
We also define $\intDtxt$ and $\intNtxt$, which have the same
meaning as $\int_k$ and $\int_{\bm{k}}$,
except that it is
understood for $\intDtxt$ that one carries out the $k^0$ integration
first, and it is understood for both $\intDtxt$ and $\intNtxt$ that 
one expands the integrand
in powers of the momenta divided by $m$.
\subsection{Vertex Correction\label{subsec:NRQCDvertex}}
Now, we calculate the NRQCD vertex correction 
to the electromagnetic current. From Eq.~(\ref{numerator}), we see that 
the vertex correction is given by
\begin{eqnarray}
\label{vertexNRQCD00}%
\Lambda^i_{\textrm{NRQCD}}
&=&
-ig_s^2C_F
\intD
\frac{1} {D_0D_1D_2}
\,
\bar{v} (p_2) \bigg\{
                      \bigg[(d-2)k^2 - 4(2 p^2 - m^2) + 8  k \cdot q
                      \bigg]
                      \gamma^i
\nonumber\\
&&\hspace{27ex}
+ 4 m k^{i} - 8 q^{i} k\!\!\!/ + 2 (2-d) k^{i} k\!\!\!/
\,\bigg\} u(p_1).
\end{eqnarray}
The vertex correction (\ref{vertexNRQCD00}) can be written as
\begin{eqnarray}
\label{LambdaNRQCD0}%
\Lambda^i_{\textrm{NRQCD}}
&=&-ig_s^2C_F \bar{v}(p_2)
\bigg\{
\Big[
(d-2)S_1-4(2p^2-m^2)S_2+8q_\mu S^\mu_3
\Big]
\gamma^i 
\nonumber\\
&&\hspace{15ex}
+2 \Big[
2mS_3^i -4\gamma_\mu S_3^\mu q^i +(2-d)\gamma_\mu S_4^{\mu i}
\Big]
\bigg\}
u(p_1),
\end{eqnarray}
where
\begin{subequations}
\label{Si-definition}%
\begin{eqnarray}
S_1 &=& \intD \frac{1}{D_1 D_2},
\\
S_2 &=& \intD \frac{1}{D_0 D_1 D_2},
\\
S^{\mu}_3 &=& \intD \frac{k^\mu}{D_0 D_1 D_2},
\\
S^{\mu\nu}_4 &=& \intD \frac{k^\mu k^\nu}{D_0 D_1 D_2}.
\end{eqnarray}
\end{subequations}
The factors in the denominator of the integrands are
defined in Eq.~(\ref{Di-definition}).
In the $Q\bar{Q}$ rest frame, the factors 
$D_i$ in Eq.~(\ref{Di-definition}) are 
\begin{subequations}
\label{poles}%
\begin{eqnarray}
D_0&=&(k^0)^2-\bm{k}^2+i\varepsilon=
(k^0-|\bm{k}|+i \varepsilon)(k^0+|\bm{k}|-i \varepsilon),
\\
D_1&=&(k^0+E)^2-\Delta^2+i\varepsilon
    =(k^0+\Delta+E-i\varepsilon)(k^0-\Delta+E+i\varepsilon),
\\
D_2&=&(k^0-E)^2-\Delta^2+i\varepsilon
    =(k^0+\Delta-E-i\varepsilon)(k^0-\Delta-E+i\varepsilon),
\end{eqnarray}
\end{subequations}
where $\Delta$ is defined by
\begin{equation}
\Delta=\sqrt{m^2+(\bm{k}+\bm{q})^2}.
\end{equation}

The following are identities that we use frequently:
\begin{subequations}
\begin{eqnarray}
\label{d2e2}%
\Delta-E&=&\frac{\bm{k}^2+2\bm{k}\cdot\bm{q}}{\Delta+E},
\\
\label{d2ke2}%
\Delta^2-(E\pm|\bm{k}|)^2&=&
\mp 2|\bm{k}|(E\mp \bm{q}\cdot \hat{\bm{k}}),
\end{eqnarray}
\end{subequations}
where $\hat{\bm{a}}=\bm{a}/|\bm{a}|$ for any spatial vector $\bm{a}$.
We first evaluate the $k^0$ integral by contour integration, closing the
contour in the upper half-plane in every case. The contributions from
the poles in the gluon, quark, and antiquark propagators are defined as
$S_{ig}$, $S_{iQ}$, and $S_{i\bar{Q}}$, respectively. Certain integrals
that we use frequently are tabulated in Appendix~\ref{appendix:IntS}.

We note that the contributions $S_{i\bar{Q}}$ correspond to the
potential region in the method of regions, and the contributions
$S_{ig}$ correspond to the soft and ultrasoft regions in the method of
regions \cite{Beneke:1997zp}. The contributions $S_{iQ}$ correspond to a
region of integration in which the temporal component of the gluon
momentum is of order $m$, but the spatial component of the gluon
momentum is of order $mv$. As we have mentioned, in the method of
regions it is assumed that this region of integration does not
contribute \cite{Beneke:1997zp}. We shall see explicitly in the
calculations that follow that this assumption is justified because the
contributions from this region of integration consist of scaleless,
power-divergent integrals, which vanish in dimensional regularization. 
In the case of a hard-cutoff regulator these contributions do not
vanish, and they must be included in the calculation of the NRQCD
corrections.
\subsubsection{${S_1}$}
The integral $S_1$ is the sum of two contributions: 
$S_1=S_{1Q}+S_{1\bar{Q}}$.

By making use of Eq.~(\ref{poles}), we 
evaluate the $k^0$ integral. The contribution from the quark pole is
\begin{equation}
\label{s1q}%
S_{1Q} =
- \frac{i}{8E}
\intN
\frac{1}{\Delta(\Delta+E)}.
\end{equation}
Expanding $1/\Delta$ and $1/(\Delta+E)$ in Eq.~(\ref{s1q})
in powers of $(\bm{k}+\bm{q})^2/m^2$,
we find that all of the terms in the
expansion are scaleless, power-divergent integrals.
Hence,
\begin{equation}
\label{s1qf}%
S_{1Q} =0.
\end{equation}

The contribution from the antiquark pole is
\begin{equation}
\label{s1qb}%
S_{1\bar{Q}} =
\frac{i}{8E}
\intN
\frac{1}{\Delta (\Delta-E-i \varepsilon)}.
\end{equation}
We use the identity (\ref{d2e2}) to reduce
the integrand in Eq.~(\ref{s1qb}) to the following form:
\begin{equation}
\label{s1qb-plus}
S_{1\bar{Q}} =
\frac{i}{8E}
\intN\left(1+\frac{E}{\Delta}\right)
\frac{1}{\bm{k}^2+2\bm{k}\cdot\bm{q}-i \varepsilon}.
\end{equation}
Expanding $1/\Delta$ in Eq.~(\ref{s1qb-plus})
in powers of $(\bm{k}+\bm{q})^2/m^2$,
we find that the expansion brings in additional factors of
$(\bm{k}+\bm{q})^2$. In each additional factor, only the term $\bm{q}^2$
survives, as the terms $\bm{k}^2+2\bm{k}\cdot\bm{q}$ lead to scaleless,
power-divergent integrals, which vanish.
As a result, we can replace $\Delta$ with $E$ in Eq.~(\ref{s1qb-plus}).
Hence, $S_{1\bar{Q}}$ is
proportional to an elementary integral $n_1$, which is defined in
Eq.~(\ref{n1}):
\begin{equation}
\label{s1qbf}%
S_{1\bar{Q}} =
\frac{i}{4E}\,n_1= - \frac{|\bm{q}|}{16 \pi E}.
\end{equation}
Using Eqs.~(\ref{delta}), (\ref{s1qf}), and (\ref{s1qbf}), we obtain
\begin{equation}
\label{s1f}%
S_1
= \frac{i}{(4\pi)^2}\,\,
i \pi \delta.
\end{equation}
\subsubsection{${S_2}$}
The integral $S_2$ is the sum of three contributions: 
$S_2=S_{2g}+S_{2Q}+S_{2\bar{Q}}$.

By making use of Eq.~(\ref{poles}), we evaluate the $k^0$ 
integral. The gluon-pole contribution is
\begin{eqnarray}
\label{s2g}%
S_{2g}&=&
-\frac{i}{2} 
\intN
\frac{1}{|\bm{k}|
[\Delta^2-(E+|\bm{k}|)^2-i\varepsilon]
[\Delta^2-(E-|\bm{k}|)^2-i\varepsilon]}
\nonumber\\
&=&
\frac{i}{8} 
\int_{\bm{k}}
\frac{1}{|\bm{k}|^3 
[E^2-(\bm{q}\cdot\hat{\bm{k}})^2]
        },
\end{eqnarray}
where we have used the identity (\ref{d2ke2}).
Making use of Eq.~(\ref{angularav2}), we find that $S_{2g}$ is proportional
to $n_0$ in Eq.~(\ref{n0}):
\begin{equation}
\label{s2g-simple}%
S_{2g}=
\frac{i}{16 E|\bm{q}|} \,n_0 
\log \left( \frac{E+|\bm{q}|}{E-|\bm{q}|} \right).
\end{equation}
Using Eqs.~(\ref{eq-delta}) and (\ref{n0}), we express $S_{2g}$
in terms of $m$ and $\delta$ as
\begin{equation}
\label{s2gf}%
S_{2g}
=
\frac{i}{32\pi^2 m^2}
  \left(
 \frac{1}{\epsilon_{\textrm{UV}}}
-\frac{1}{\epsilon_{\textrm{IR}}}\right)
  \frac{1-\delta^2}{2\delta}
  \log\left(\frac{1+\delta}{1-\delta}\right).
\end{equation}

The contribution from the quark pole is
\begin{eqnarray}
\label{s2q}%
S_{2Q}&=&
- \frac{i}{8E}
\intN
\frac{1}{\Delta(\Delta+E)
        [(\Delta+E)^2-\bm{k}^2+i \varepsilon]}
\nonumber\\
&=&
- \frac{i}{8E}
\sum_{n=0}^\infty
\intN
\frac{\bm{k}^{2n}}{\Delta(\Delta+E)^{2n+3}}.
\end{eqnarray}
Now we expand $1/\Delta$ and $1/(\Delta+E)$ in Eq.~(\ref{s2q}) 
in powers of $(\bm{k}+\bm{q})^2/m^2$. 
All of the terms in the expansion yield
scaleless, power-divergent integrals, which vanish.
Therefore, we have
\begin{equation}
\label{s2qf}%
S_{2Q}=0.
\end{equation}

The contribution from the antiquark pole is
\begin{equation}
\label{s2qb}%
S_{2\bar{Q}}=
-\frac{i}{8E}
\intN
\frac{1}{\Delta(\Delta-E-i \varepsilon)
        [\bm{k}^2-(\Delta-E)^2-i\varepsilon ]}.
\end{equation}
If we use the relation (\ref{d2e2}), we obtain
\begin{equation}
S_{2\bar{Q}}=
-\frac{i}{8E}
\intN
\left(1+\frac{E}{\Delta}\right)
\frac{1}{\bm{k}^2(\bm{k}^2+2\bm{k}\cdot\bm{q}-i \varepsilon)
        \left[1-\frac{1}{\bm{k}^2}
                \left(\frac{\bm{k}^2+2\bm{k}\cdot\bm{q}}
               {\Delta+E}
          \right)^2
        \right]}.
\end{equation}
The denominator factor in the brackets can be expanded to give
\begin{equation}
\label{s2qb2-old}%
S_{2\bar{Q}}
=-\frac{i}{8E}
\intN
\left(1+\frac{E}{\Delta}\right)
\left[
\frac{1}
      {\bm{k}^{2}(\bm{k}^2+2\bm{k}\cdot\bm{q}-i \varepsilon)}
+
\sum_{n=1}^\infty
\frac{(\bm{k}^2+2\bm{k}\cdot\bm{q})^{2n-1}}
      {\bm{k}^{2n+2}(\Delta + E)^{2n}}
\right].
\end{equation}
Now we expand $E/\Delta$ and $1/(\Delta + E)$ in powers of 
$(\bm{k}+\bm{q})^2/m^2$. The expansion brings in additional 
factors of $(\bm{k}+\bm{q})^2$ in each term in the integrand of
Eq.~(\ref{s2qb2-old}). In each additional factor ($\bm{k}+\bm{q})^2$, only 
the term $\bm{q}^2$ survives, as the terms $\bm{k}^2+2\bm{k}\cdot\bm{q}$
lead to scaleless, power-divergent integrals. 
Therefore, we can 
replace $\Delta$  in Eq.~(\ref{s2qb2-old}) with $E$. Furthermore, in the 
numerator of the second term in brackets in  Eq.~(\ref{s2qb2-old}), only the 
term $(2\bm{k}\cdot \bm{q})^{2n-1}$ survives, as the other terms lead to 
scaleless, power-divergent integrals. Then, we have
\begin{equation}
\label{s2qb2}%
S_{2\bar{Q}}
=-\frac{i}{4E}
\intN
\left[
\frac{1}
      {\bm{k}^{2}(\bm{k}^2+2\bm{k}\cdot\bm{q}-i \varepsilon)}
+
\frac{1}{2}
\sum_{n=1}^\infty
\frac{(\bm{k}\cdot\bm{q})^{2n-1}}
      {\bm{k}^{2n+2} E^{2n}}
\right].
\end{equation}
The term proportional to $(\bm{k}\cdot\bm{q})^{2n-1}$ yields a 
scaleless, logarithmically divergent integral.
However, this integral vanishes because
the integrand is an odd function of $\bm{k}$. Thus, only
the first term in the brackets in Eq.~(\ref{s2qb2}) survives, and
we find that
\begin{equation}
\label{s2qbf}%
S_{2\bar{Q}}=
-\frac{i}{4E}n_2
=
-\frac{1}{64\pi E|\bm{q}|}\left(
\frac{1}{\epsilon_{\textrm{IR}}}
+\log 
\frac{\pi \mu^2 e^{- \gamma_{_{\!\textrm{E}}}}  }
{\bm{q}^2} 
+i\pi
\right),
\end{equation}
where $n_2$ is defined in Eq.~(\ref{n2}).

Making use of Eqs.~(\ref{eq-delta}), (\ref{s2gf}), (\ref{s2qf}),
and (\ref{s2qbf}), we obtain
\begin{equation}
\label{s2f}%
S_2
= \frac{i}{(4\pi)^2}\,\,
\frac{1-\delta^2}{4m^2}
\Bigg[
 2 L(\delta)
\left(
\frac{1}{\epsilon_{\textrm{UV}}}
-\frac{1}{\epsilon_{\textrm{IR}}}
\right)
-\frac{\pi^2}{\delta}
+ \frac{i \pi}{\delta}
\left(
\frac{1}{\epsilon_{\textrm{IR}}}
+ \log \frac{\pi \mu^2 e^{-\gamma_{_{\!\textrm{E}}}}  }
                 {\bm{q}^2 }
\right)
\Bigg],
\end{equation}
where $L(\delta)$ is defined in Eq.~(\ref{L-delta}).
\subsubsection{$S_3^\mu$}
The integral $S_3^\mu$ is the sum of three contributions: 
$S_3^\mu=S_{3g}^\mu+S_{3Q}^\mu+S_{3\bar{Q}}^\mu$.

We first evaluate $S_3^0$.
The integral of $S^{0}_3$ over $k^0$ is identical to the integral of $S_2$
over $k^0$ except that, in $S_3^0$, the result contains an additional factor
of $k^0$ evaluated at the gluon, quark, or antiquark pole.
Thus, by making use of
Eqs.~(\ref{d2e2}), (\ref{s2g}), (\ref{s2q}), and (\ref{s2qb2-old}),
we  obtain
\begin{subequations}
\label{s30}%
\begin{eqnarray}
\label{s3g0f}%
S_{3g}^0&=&
-\frac{i}{8}
\int_{\bm{k}}
\frac{1}{\bm{k}^2[ E^2-(\bm{q}\cdot\hat{\bm{k}})^2]},
\\
\label{s3q0}%
S^0_{3Q}&=&
\frac{i}{8E}
\sum_{n=0}^{\infty}
\intN
\frac{\bm{k}^{2n}}{\Delta(\Delta+E)^{2n+2}},
\\
\label{s3qb0}%
S^0_{3\bar{Q}}&=&
\frac{i}{8E}
\sum_{n=0}^\infty
\intN
\frac{(\bm{k}^2+2\bm{k}\cdot\bm{q})^{2n}}
      {\bm{k}^{2n+2}\Delta(\Delta+E)^{2n}}.
\end{eqnarray}
\end{subequations}
$S^0_{3g}$ is a scaleless, power-divergent integral, which vanishes.
In $S^0_{3Q}$ and $S^0_{3\bar{Q}}$ we expand $1/\Delta$ and 
$1/(\Delta+E)$ in powers of $(\bm{k}+\bm{q})^2/m^2$.
We find that every term in the expansions leads to a
scaleless, power-divergent integral, which vanishes.
Hence,
\begin{equation}
\label{s30f}%
S^0_{3}=0.
\end{equation}

Next we compute the spatial components $S^{i}_3$.
The integral of $S^{i}_3$ over $k^0$ is identical to the integral of $S_2$
over $k^0$ except that, in $S^{i}_3$, the result contains an additional factor
of $k^i$.  Thus, by making use of
Eqs.~(\ref{s2g}) and (\ref{s2q}), we  obtain
\begin{subequations}
\label{s3v}%
\begin{eqnarray}
\label{s3gv}%
S^{i}_{3g}&=&
\frac{i}{8}
\int_{\bm{k}}
\frac{{k}^i}
     {|\bm{k}|^3
      [ E^2-(\bm{q}\cdot \hat{\bm{k}})^2]
    },
\\
\label{s3qv}%
S^{i}_{3Q}&=&
- \frac{i}{8E}
\sum_{n=0}^\infty
\intN
\frac{{k}^i\bm{k}^{2n}}{\Delta(\Delta+E)^{2n+3}}.
\end{eqnarray}
\end{subequations}
$S^i_{3g}$ is a scaleless, power-divergent integral, which vanishes.
Expanding $1/\Delta$ and $1/(\Delta+E)$ in $S^{i}_{3Q}$ in powers of 
$(\bm{k}+\bm{q})^2/m^2$,
we also obtain only scaleless, power-divergent integrals, which vanish.
If we multiply the second term in brackets in Eq.~(\ref{s2qb2}) by
$k^i$, we obtain only scaleless, power-divergent integrals.  Hence, 
\begin{equation}
\label{s3qbv}%
S^{i}_{3 \bar{Q}}=
- \frac{i}{4 E} \int_{\bm{k}}
\frac{{k}^i}{\bm{k}^2 (\bm{k}^2 + 2 \bm{k}\cdot\bm{q} - i \varepsilon)}.
\end{equation}
After making a standard reduction of the tensor integral in 
Eq.~(\ref{s3qbv}) to a scalar
integral, we obtain
\begin{equation}
\label{s3vqbf}%
S^{i}_{3 \bar{Q}}=
- \frac{i\,{q}^i}{8 E\bm{q}^2} 
\int_{\bm{k}}
\frac{(\bm{k}^2 + 2 \bm{k}\cdot\bm{q})-\bm{k}^2}
     {\bm{k}^2 (\bm{k}^2 + 2 \bm{k}\cdot\bm{q} - i \varepsilon)}
=
\frac{i\,{q}^i}{8 E\bm{q}^2} n_1
=
- \frac{1}{32 \pi} \frac{{q}^i}{E|\bm{q}|},
\end{equation}
where $n_1$ is defined in Eq.~(\ref{n1}) and we have discarded
scaleless, power-divergent integrals. Hence,
\begin{equation}
\label{s3vf}%
S^{i}_{3}=
- \frac{1}{32 \pi} \frac{{q}^i}{E|\bm{q}|}.
\end{equation}

Writing our results in Eqs.~(\ref{s30f}) and (\ref{s3vf}) in 
covariant form, we obtain 
\begin{equation}
\label{s3muf}%
S^\mu_3 
= \frac{i}{(4\pi)^2}\,
\frac{1-\delta^2}{2m^2}\,
\frac{i \pi}{\delta}\, q^\mu,
\end{equation}
where we have made use of Eq.~(\ref{eq-delta}) to
express $E$ and $|\bm{q}|$ in terms of $\delta$.
\subsubsection{$S_4^{\mu\nu}$}
The integral $S_4^{\mu\nu}$ is the sum of three contributions: 
$S_4^{\mu \nu} =S_{4g}^{\mu \nu}+S_{4Q}^{\mu \nu}+S_{4\bar{Q}}^{\mu \nu}$. 

We first evaluate $S_4^{00}$.
The integral of $S^{00}_4$ over $k^0$ is identical to the integral of $S_3^0$
over $k^0$ except that, in $S_4^{00}$, 
the result contains an additional factor
of $k^0$ evaluated at the gluon, quark, or antiquark pole.
Thus, by making use of
Eqs.~(\ref{d2e2}) and (\ref{s30}), we  obtain
\begin{subequations}
\label{s004}%
\begin{eqnarray}
S^{00}_{4g}&=&
\frac{i}{8}
\int_{\bm{k}} 
\frac{1}
{|\bm{k}| [E^2-(\bm{q}\cdot \hat{\bm{k}})^2]},
\\
S^{00}_{4Q}&=&
-\frac{i}{8E}
\sum_{n=0}^{\infty}
\intN
\frac{\bm{k}^{2n}}{\Delta(\Delta+E)^{2n+1}},
\\
S^{00}_{4\bar{Q}}&=&
-\frac{i}{8E}
\sum_{n=0}^\infty
\intN\frac{(\bm{k}^2+2 \bm{k}\cdot\bm{q} )^{2n+1}}
                  {\bm{k}^{2n+2}\Delta(\Delta+E)^{2n+1}} .
\end{eqnarray}
\end{subequations}
Every integral in Eq.~(\ref{s004}) is a scaleless, 
power-divergent integral.
Hence,
\begin{equation}
\label{s004f}%
S^{00}_{4}=0.
\end{equation}

Next we compute ${S}_4^{0i}$.
The integral of ${S}_4^{0i}$ over $k^0$ is identical to the integral of $S_3^0$
over $k^0$ except that, in ${S}_4^{0i}$, 
the result contains an additional factor
of $k^i$.  Thus, by making use of Eq.~(\ref{s30}), we  obtain
\begin{subequations}
\label{s0i4}%
\begin{eqnarray}
{S}^{0i}_{4g}&=&
- \frac{i}{8}
\int_{\bm{k}}\frac{k^i}
{\bm{k}^{2}[
 E^2- (\bm{q}\cdot \hat{\bm{k}})^2
                ]
    },
\\
{S}^{0i}_{4Q}&=&
\frac{i}{8E}
\sum_{n=0}^\infty
\intN
\frac{k^i \bm{k}^{2n}}{\Delta(\Delta+E)^{2n+2}},
\\
{S}^{0i}_{4 \bar{Q}}&=&
\frac{i}{8E}
\sum_{n=0}^\infty
\intN
\frac{k^i(\bm{k}^2+2\bm{k}\cdot\bm{q})^{2n}}
      {\bm{k}^{2n+2}\Delta(\Delta+E)^{2n}}.
\end{eqnarray}
\end{subequations}
$S^{0i}_{4g}$ is a scaleless, power-divergent integral, which vanishes.
$S^{0i}_{4Q}$ and $S^{0i}_{4\bar{Q}}$ also vanish, once we expand $1/\Delta$ 
and $1/(\Delta+E)$ in powers of $(\bm{k}+\bm{q})^2/m^2$. Thus,
\begin{equation}
\label{s0i4f}%
S^{0i}_{4}=0.
\end{equation}

Finally, we evaluate the integrals $S^{ij}_4$. The integral of ${S}_4^{ij}$
over $k^0$ is identical to the integral of $S_3^{i}$ over $k^0$ except that,
in ${S}_4^{ij}$, the result contains an additional factor of $k^j$.
Thus, by making use of Eqs.~(\ref{s3v}) and (\ref{s3qbv}), we  obtain
\begin{subequations}
\label{sij4}%
\begin{eqnarray}
{S}^{ij}_{4g}&=&
\frac{i}{8}
\int_{\bm{k}}
\frac{k^i k^j}
     {|\bm{k}|^3[
 E^2- (\bm{q}\cdot \hat{\bm{k}})^2
                ]
    },
\\
{S}^{ij}_{4Q}&=&
- \frac{i}{8E}
\sum_{n=0}^\infty
\intN
\frac{k^i k^j \bm{k}^{2n}}{\Delta(\Delta+E)^{2n+3}},
\\
\label{sijqb4-simple}%
{S}^{ij}_{4 \bar{Q}}&=&
-\frac{i}{4E}
\int_{\bm{k}}
\frac{k^i k^j}
      {\bm{k}^2(\bm{k}^2+2\bm{k}\cdot\bm{q}-i \varepsilon)}.
\end{eqnarray}
\end{subequations}
$S^{ij}_{4g}$ is a scaleless, power-divergent integral, which vanishes.
${S}^{ij}_{4Q}$ also vanishes, once we expand $1/\Delta$ and $1/(\Delta+E)$
in powers of $(\bm{k}+\bm{q})^2/m^2$. The tensor integral 
${S}^{ij}_{4 \bar{Q}}$ in Eq.~(\ref{sijqb4-simple}) must be a linear
combination of the two symmetric tensors $\delta^{ij}$ and 
${q}^i{q}^j$. By contracting these tensors into Eq.~(\ref{sijqb4-simple}),
we determine the coefficients of the linear combination. The result is
\begin{eqnarray}
\label{sijqb4-n1}%
{S}^{ij}_{4\bar{Q}}
&=&
-\frac{i}{4E(d-2)}
\left[ 
\delta^{ij}\left( n_1 -\frac{1}{4 \bm{q}^2}n_3 \right)
-\frac{q^i q^j}{\bm{q}^2}\left(n_1-\frac{d-1}{4\bm{q}^2} n_3 \right)
\right]
\nonumber \\
&=&
\frac{|\bm{q}|}{32\pi(d-2)E}
\bigg[
\delta^{ij}
+(d-3)\frac{q^i q^j}{\bm{q}^2}
\bigg],
\end{eqnarray}
where $n_3$ is defined in Eq.~(\ref{n3}). Because $S_{4 g}^{ij}$ and
$S_{4 \bar{Q}}^{ij}$ vanish, we find that $S_4^{ij} = S_{4 \bar{Q}}^{ij}$.
The integral in Eq.~(\ref{sijqb4-n1}) is finite and, therefore,
we may set $d=4$.

The covariant form of the integral $S^{\mu\nu}_4$ at $d=4$ is then 
\begin{equation}
\label{s4f}%
S_4^{\mu \nu}
= \frac{i}{(4\pi)^2}\,\,
\frac{ i\pi\delta}{4}
\bigg[
 g^{\mu \nu}
 -  \frac{1-\delta^2}{m^2}
  \left(p^\mu p^\nu+\frac{q^\mu q^\nu}{\delta^2}
  \right)
\bigg].
\end{equation}
\subsubsection{Summary of the NRQCD vertex correction}
Substituting
$S_1$ -- $S_4^{\mu \nu}$ in Eqs.~(\ref{s1f}), (\ref{s2f}), (\ref{s3muf}), and
(\ref{s4f}) into Eq.~(\ref{LambdaNRQCD0}) and using the equations of motion
in Eq.~(\ref{diraceq}), we obtain
\begin{subequations}
\label{LambdaNRQCDf}%
\begin{eqnarray}
\Lambda_{\textrm{NRQCD}}
&=&
\frac{\alpha_s C_F}{4 \pi}
(1+\delta^2)
\bigg[
2 L(\delta)
\left(
\frac{1}{\epsilon_{\textrm{IR}}}
-\frac{1}{\epsilon_{\textrm{UV}}}
\right)
+\frac{\pi^2}{\delta}
- \frac{i \pi}{\delta}
 \left(
 \frac{1}{\epsilon_{\textrm{IR}}}
+ \log \frac{\pi \mu^2e^{-\gamma_{_{\!\textrm{E}}}}  }
                 {\bm{q}^2 }
+
\frac{3 \delta^2 }{1+\delta^2} \right)
\bigg],
\nonumber\\
&&
\\
H_{\textrm{NRQCD}}
 &=&
\frac{\alpha_s C_F}{4\pi} \,\frac{1-\delta^2}{m}
\left( - \frac{i\pi}{\delta}\right) .
\end{eqnarray}
\end{subequations}
\subsection{Wave-function Renormalization\label{subsec:NRQCDZQ}}
In the Feynman gauge, the self energy of the heavy quark, evaluated
at four-momentum $p_1$, is
\begin{equation}
[\Sigma(p_1)]_{\textrm{NRQCD}}=
-ig_s^2C_F\intD
\frac{\gamma_\mu (/\!\!\!p_1+/\!\!\!k+m)\gamma^\mu }
{(k^2+i\varepsilon)[(p_1+k)^2-m^2+i\varepsilon]},
\end{equation}
where $m$ is the mass of the heavy quark and  $k$ is the loop momentum,
which has been chosen to be the momentum of the virtual gluon.
In $d$ dimensions, we find that the numerator factor reduces to
\begin{equation}
\label{sigmap1}%
[\Sigma(p_1)]_{\textrm{NRQCD}}=
-ig_s^2C_F\intD
\frac{(2-d)(/\!\!\!p_1+/\!\!\!k)+dm}
{(k^2+i\varepsilon)[(p_1+k)^2-m^2+i\varepsilon]}.
\end{equation}

The heavy-quark wave-function renormalization $Z_Q$ is defined by
\begin{eqnarray}
\label{zq-as}%
[Z_Q]_{\textrm{NRQCD}}
&=&\bigg[
1-
\left.
\frac{p_1^\mu}{m}
\frac{\partial[\Sigma(p_1)]_{\textrm{NRQCD}}}{\partial p_1^\mu}
\right|_{/\!\!\!p_1=m}
\bigg]^{-1}
\nonumber\\
&=&
1+
\left.
\frac{p_1^\mu}{m}
\frac{\partial[\Sigma(p_1)]_{\textrm{NRQCD}}}{\partial p_1^\mu}
\right|_{/\!\!\!p_1=m}+O(\alpha_s^2).
\end{eqnarray}

Differentiating Eq.~(\ref{sigmap1}), we find that
\begin{eqnarray}
\label{dsdp}%
\left.
\frac{p_1^\mu}{m}
\frac{\partial[\Sigma(p_1)]_{\textrm{NRQCD}}}{\partial p_1^\mu}
\right|_{/\!\!\!p_1=m}
&=&
-ig_s^2C_F\intD
\bigg\{
\frac{2-d}
{D_0D_1}
-
\frac{2[(2-d)(/\!\!\!k+m)+dm](p_1\cdot k+m^2)}
{mD_0D_1^2}
\bigg\}
\nonumber\\
&=&
-ig_s^2C_F\intD
\bigg[
\frac{2-d}
{D_0D_1}
-
\frac{(2-d)/\!\!\!k+2m}{m}
\left(
\frac{1}{D_0D_1}
-
\frac{1}{D_1^2}
+
\frac{2m^2}{D_0D_1^2}
\right)
\bigg],
\nonumber\\
\end{eqnarray}
where $D_0$ and $D_1$ are defined in Eq.~(\ref{Di-definition}).
The expression in Eq.~(\ref{dsdp}) can be written in terms of 
the integrals $T_{02}$, $T_{11}$, $T_{12}$, $T_{02}^\mu$, $T_{11}^\mu$, 
and $T_{12}^{\mu}$,
which are defined by
\begin{subequations}
\label{ti-definition}%
\begin{eqnarray}
T_{ab}
&=&
\intD\frac{1}{D_0^a D_1^b},
\\
T_{ab}^{\mu}
&=&
\intD\frac{k^\mu}{D_0^a D_1^b}.
\end{eqnarray}
\end{subequations}
These integrals are evaluated in Appendix~\ref{appendix:T-integrals},
and the results are summarized in Eqs.~(\ref{t-vanishing}) and (\ref{t12f}).
The only nonvanishing integral is $T_{12}$. Hence, 
\begin{eqnarray}
\label{dsdpf}%
\left.
\frac{p_1^\mu}{m}
\frac{\partial[\Sigma(p_1)]_{\textrm{NRQCD}}}{\partial p_1^\mu}
\right|
_{/\!\!\!p_1=m}
=
4ig_s^2C_F\, m^2\, T_{12}.
\end{eqnarray}
Making use of Eqs.~(\ref{zq-as}),
(\ref{dsdpf}), and (\ref{t12f}),
we obtain the heavy-quark wave-function renormalization in NRQCD:
\begin{equation}
\label{ZQNRQCD}%
[Z_Q]_{\textrm{NRQCD}}
= 1+ \frac{\alpha_s C_F}{2 \pi} \left( 
  \frac{1}{\epsilon_{\textrm{UV}}}
- \frac{1}{\epsilon_{\textrm{IR}}}
\right)+O(\alpha_s^2).
\end{equation}
\subsection{Summary of NRQCD results\label{subsec:NRQCDsummary}}
Making use of Eqs.~(\ref{LambdaNRQCDf}) and (\ref{ZQNRQCD}),
we find that
\begin{subequations}
\label{GHNRQCDf}%
\begin{eqnarray}
G_{\textrm{NRQCD}}
&=& 
1+ \frac{\alpha_s C_F}{4 \pi} 
\left\{ 
2 [(1+\delta^2) L(\delta) - 1]
\Big( 
\frac{1}{\epsilon_{\textrm{IR}}}
-\frac{1}{\epsilon_{\textrm{UV}}}
\Big)
\right.
\nonumber \\
&& \hspace{8ex}
\left.
+
(1+\delta^2)
\left[
\frac{\pi^2}{\delta} 
- \frac{i \pi}{\delta} 
\left( 
\frac{1}{\epsilon_{\textrm{IR}}} 
+ \log \frac{\pi \mu^2 e^{-\gamma_{_{\!\textrm{E}}}}}
{\bm{q}^2} 
+ \frac{3 \delta^2}{1+\delta^2}
\right)
\right]
\right\},
\\
H_{\textrm{NRQCD}}
 &=& 
\frac{\alpha_s C_F}{4\pi} \,\frac{1-\delta^2}{m}
\left( - \frac{i\pi}{\delta}\right) .
\end{eqnarray}
\end{subequations}

Expanding Eq.~(\ref{amp-A-B-NRQCD}) through order $v^2$,
we obtain
\begin{eqnarray}
\label{Ai-NRQCD}%
i\left[{\cal A}_{Q\bar Q_1}^{i}\right]_{\rm NRQCD} 
&=&
\eta^\dagger \sigma^i \xi 
\Bigg[
\,
1+ \frac{\alpha_s C_F}{4 \pi} 
\bigg\{
\frac{8 v^2}{3}
\left(
\frac{1}{\epsilon_{\textrm{IR}}}
-\frac{1}{\epsilon_{\textrm{UV}}}
\right)
\nonumber \\
&&\hspace{10ex}+
\left( 1 + \frac{3 v^2}{2} \right)
\bigg[ \frac{\pi^2}{v} - \frac{i \pi}{v}
\bigg(
  \frac{1}{\epsilon_{\textrm{IR}}}
+\log
   \frac{\pi \mu^2
             e^{-\gamma_{_{\!\textrm{E}}}}
            }{\bm{q}^2}
\bigg)
\bigg]
-3 i \pi v
\bigg\}
\,
\Bigg]
\nonumber \\
&-& 
\frac{q^i\eta^\dagger \bm{q}\cdot\bm{\sigma} \xi}{2 m^2}
\left\{ 
1+ \frac{\alpha_s C_F}{4 \pi} 
\left[
 \frac{\pi^2}{v} 
- 
\frac{i \pi}{v}
\left(
\frac{1}{\epsilon_{\textrm{IR}}} 
+\log
\frac{\pi \mu^2 e^{-\gamma_{_{\!\textrm{E}}}}}
{\bm{q}^2} 
\right)
-
\frac{2 i \pi}{v}
\right]
\right\}+ O(v^3).
\nonumber\\
\end{eqnarray}
Comparing Eq.~(\ref{Ai-NRQCD}) with Eqs.~(4.28) and (4.29) of 
Ref.~\cite{Luke:1997ys}, we find agreement. We have also checked 
Eq.~(\ref{Ai-NRQCD}) by carrying out a conventional calculation in NRQCD.
\section{Results for the short-distance coefficients\label{sec:RESCoeff}}
Now we can collect the results of our calculations and obtain the 
short-distance coefficients. By making use of Eqs.~(\ref{DeltaAB}), 
(\ref{QCDAB}), and (\ref{GHNRQCDf}), we find that
\begin{subequations}
\label{DeltaGH}%
\begin{eqnarray}
\Delta G^{(1)}  
&=& \frac{\alpha_s C_F}{4 \pi}
\bigg\{ 
2 \,\big[ (1+\delta^2) L(\delta) - 1\big]
\left(
          \frac{1}{\epsilon_{\textrm{UV}}}
       + \log \frac{4\pi \mu^2e^{-\gamma_{_{\!\textrm{E}}}}}{m^2} 
\right)
\nonumber \\
&& \hspace{8ex} 
+\,6\delta^2 L(\delta) - 4 (1+\delta^2) K (\delta) - 4 \bigg\},
\\
\label{DeltaH}%
\Delta H^{(1)}  &=& 
\frac{\alpha_s C_F}{4 \pi} \frac{2 (1-\delta^2)}{m} L(\delta).
\end{eqnarray}
\end{subequations}
As expected, the infrared poles in $G^{(1)}$ and $G_{\rm NRQCD}^{(1)}$
have canceled in $\Delta G^{(1)}$. Note that
$\Delta G^{(1)}$ and $\Delta H^{(1)}$ are real and contain only even
powers of $v=|\bm{q}|/m$. Renormalizing the matrix elements in the 
$\overline{\rm MS}$ scheme, we obtain
\begin{equation}
\label{DeltaGMS}%
\Delta G^{(1)}_{\overline{\rm MS}} 
= \frac{\alpha_s C_F}{4 \pi}
\left\{ 
2\, \big[ (1+\delta^2) L(\delta) - 1\big]
\log \frac{\mu^2}{m^2} 
+
6 \delta^2 L(\delta) - 4 (1+\delta^2) K (\delta) - 4 \right\},
\end{equation}
where now $\mu$ is the NRQCD factorization scale. Using Eq.~(\ref{an0-bn0}), 
we obtain the short-distance coefficients $a_n^{(0)}$ and $b_n^{(0)}$:
\begin{subequations}
\begin{eqnarray}
\label{lo-results-a}%
a_n^{(0)}&=&\delta_{n0},
\\
\label{lo-results-b}%
b_1^{(0)}&=&  - \frac{1}{2 m^2}
,
\\
\label{lo-results-c}%
b_2^{(0)}&=&    \frac{3}{8 m^4}
,
\\
\label{lo-results-d}%
b_3^{(0)}&=&  - \frac{5}{16 m^6} .
\end{eqnarray}
\end{subequations}
The results in Eqs.~(\ref{lo-results-a})--(\ref{lo-results-c}) agree 
with those in Eq.~(5.5) of Ref.~\cite{Bodwin:2002hg} and those in 
Eqs.~(3.13)--(3.20) of Ref.~\cite{Brambilla:2006ph}.
Using Eqs.~(\ref{ab-MSbar}), (\ref{DeltaH}), and (\ref{DeltaGMS}),
we obtain the short-distance coefficients 
$\left[a_n^{(1)}\right]_{\overline{\rm MS}}$ and 
$\left[b_n^{(1)}\right]_{\overline{\rm MS}}$:
\begin{subequations}
\begin{eqnarray}
\left[a_0^{(1)}\right]_{\overline{\rm MS}}&=&
\frac{\alpha_s C_F}{4 \pi}\,\,
(- 8),
\\
\left[a_1^{(1)}\right]_{\overline{\rm MS}}&=&
\frac{\alpha_s C_F}{4 \pi}\,\, 
\frac{1}{m^2}
\left( \frac{2}{9} + \frac{8}{3} \log \frac{\mu^2}{m^2} \right),
\\
\left[a_2^{(1)}\right]_{\overline{\rm MS}}&=&
\frac{\alpha_s C_F}{4 \pi}\,\, 
\frac{1}{m^4} 
\left( -\frac{92}{75} - \frac{8}{5} \log \frac{\mu^2}{m^2} \right),
\\
\left[a_3^{(1)}\right]_{\overline{\rm MS}}&=&
\frac{\alpha_s C_F}{4 \pi}\,\, 
\frac{1}{m^6} 
\left( \frac{13744}{11025} + \frac{128}{105} \log \frac{\mu^2}{m^2} \right),
\\
\left[b_1^{(1)}\right]_{\overline{\rm MS}}&=&
\frac{\alpha_s C_F}{4 \pi}\,\,
\frac{2}{m^2},
\\
\left[b_2^{(1)}\right]_{\overline{\rm MS}}&=&
-
\frac{\alpha_s C_F}{4 \pi}\,\, 
\frac{1}{ m^4 }
\left( \frac{7}{9} + \frac{4}{3} \log \frac{\mu^2}{m^2} \right),
\\
\left[b_3^{(1)}\right]_{\overline{\rm MS}}&=&
\frac{\alpha_s C_F}{4 \pi}\,\, 
\frac{1}{ m^6 } 
\left( \frac{107}{150} + \frac{9}{5} 
\log \frac{\mu^2}{m^2} \right).
\end{eqnarray}
\end{subequations}
The operators ${\cal O}_{A0}$, ${\cal O}_{A1}$, and ${\cal O}_{B1}$ in
Eq.~(\ref{OPerators}) correspond to the operators that were considered
in Ref.~\cite{Luke:1997ys}, provided that one neglects the gauge fields
in the latter operators.
Therefore, short-distance coefficients 
$\left[a_0\right]_{\overline{\rm MS}}$,
$\left[a_1\right]_{\overline{\rm MS}}$, and
$\left[b_1\right]_{\overline{\rm MS}}$ are related to
the coefficients $c_i$ in Eq.~(4.29) of Ref.~\cite{Luke:1997ys} as follows:
\begin{subequations}
\begin{eqnarray}
\left[a_0\right]_{\overline{\rm MS}}&=& c_1,
\\
\left[a_1\right]_{\overline{\rm MS}}&=& -\frac{1}{m^2}\,c_3,
\\
\left[b_1\right]_{\overline{\rm MS}}&=& -\frac{1}{2 m^2}\,c_2.
\end{eqnarray}
\end{subequations}
Our results for these short-distance coefficients agree with those in
Eq.~(4.29) of Ref.~\cite{Luke:1997ys}.

\subsection{Resummation}
Let us define ratios of the $S$-wave $Q\bar Q$ operator matrix elements 
to the $S$-wave $Q\bar Q$ operator matrix element of lowest order in $v$:
\begin{eqnarray}
\label{me-ratios}%
\langle \bm{q}^{2n}\rangle_{H({}^3S_1)}&=&
\frac{\langle 0|\mathcal{O}_{An}^i|H({}^3S_1)\rangle}
     {\langle 0|\mathcal{O}^{i}_{A0}|H({}^3S_1)\rangle},
\end{eqnarray}
where $\mathcal{O}^{i}_{An}$ is defined in Eq.~(\ref{OPeratorsA}),
and we have used the property that the ratios are independent of the value
of the index $i$. In Ref.~\cite{Bodwin:2006dn}, it was shown that these
ratios of operator matrix elements are related according to a
generalized Gremm-Kapustin relation \cite{Gremm:1997dq}:
\begin{equation}
\label{gen-Gremm-Kapustin}%
\left[\langle \bm{q}^{2n}\rangle_{H({}^3S_1)}\right]_{\overline{\rm MS}}=
\left[\langle \bm{q}^2\rangle_{H({}^3S_1)}\right]^n_{\overline{\rm MS}}.
\end{equation}
This relation holds for the matrix elements in spin-independent-potential
models. Hence, for each value of $n$, it  holds up to 
corrections of relative order $v^2$. 

We can use the relation (\ref{gen-Gremm-Kapustin}) to resum a class
of relativistic corrections to the quarkonium electromagnetic current.  From 
Eqs.~(\ref{coeffs-s}) and (\ref{ab-MSbar}), 
we find that
\begin{eqnarray}
\label{resummed}%
&& \hspace{-10ex}
\sum_{n=0}^\infty
\left(s_n^{(0)} +  \left[s_n^{(1)}\right]_{\overline{\rm MS}}
\right) 
\langle 0|\mathcal{O}^{i}_{An}|H({}^3S_1)\rangle
\nonumber \\
&=& 
\left.
\left\{
\left[ 1- \frac{\bm{q}^2}{E (E+m) (d-1)} \right] 
\left(1+\Delta G^{(1)}_{\overline{\textrm{MS}}}\right)
- \frac{\bm{q}^2}{E (d-1)} 
\Delta H^{(1)}
\right\}\right|_{\bm{q}^2=\langle \bm{q}^2 \rangle_{H({}^3S_1)}}
\nonumber\\[1.5ex]
&&
\hspace{0ex}
\times\,
\langle 0|\mathcal{O}^{i}_{A0}|H({}^3S_1)\rangle.
\end{eqnarray}
Because the relation (\ref{gen-Gremm-Kapustin}) contains corrections of
relative order $v^2$ at each order $v^{2n}$, the resummation in
Eq.~(\ref{resummed}) does not improve the nominal accuracy beyond order
$v^4$. The resummation might, however, improve the numerical accuracy
beyond the accuracy that is obtained through order $v^4$ if the
coefficients in the velocity expansion grow rapidly with the order in
$v$. In any case, it is interesting to use the resummed result to
examine the rate of convergence of the velocity expansion.

\subsection{Numerical results and convergence of the velocity expansion}

Let us evaluate the sums of products of $S$-wave short-distance
coefficients and operator matrix elements, using the relation
(\ref{gen-Gremm-Kapustin}). For $\langle\bm{q}^{2}\rangle_{H({}^3S_1)}$, 
we take the central value of the
$J/\psi$ matrix element from Ref.~\cite{Bodwin:2007fz}:
$\langle\bm{q}^{2}\rangle_{J/\psi}=0.441\,\textrm{GeV}^2$. 
Taking $m_c=1.5~\,\textrm{GeV}$ and setting $\mu=m_c$, we find that
\begin{subequations}
\label{num-coefficient}%
\begin{eqnarray}
\sum_{n=0}^0\left[s_n^{(1)}\right]_{\overline{\rm MS}}
\,
[\langle\bm{q}^{2}\rangle_{J/\psi}]^{n}_{\overline{\rm MS}}
&=&
- \frac{\alpha_s C_F}{4 \pi}\times
8
,
\\
\sum_{n=0}^1\left[s_n^{(1)}\right]_{\overline{\rm MS}}
\,
[\langle\bm{q}^{2}\rangle_{J/\psi}]^{n}_{\overline{\rm MS}}
&=&
- \frac{\alpha_s C_F}{4 \pi}\times
7.826
,
\\
\sum_{n=0}^2\left[s_n^{(1)}\right]_{\overline{\rm MS}}
\,
[\langle\bm{q}^{2}\rangle_{J/\psi}]^{n}_{\overline{\rm MS}}
&=&
- \frac{\alpha_s C_F}{4 \pi}\times
7.883
,
\\
\sum_{n=0}^3\left[s_n^{(1)}\right]_{\overline{\rm MS}}
\,
[\langle\bm{q}^{2}\rangle_{J/\psi}]^{n}_{\overline{\rm MS}}
&=&
- \frac{\alpha_s C_F}{4 \pi}\times
7.872
,
\\
\sum_{n=0}^\infty\left[s_n^{(1)}\right]_{\overline{\rm MS}}
\,
[\langle\bm{q}^{2}\rangle_{J/\psi}]^{n}_{\overline{\rm MS}}
&=&
- \frac{\alpha_s C_F}{4 \pi}\times
7.873 .
\end{eqnarray}
\end{subequations}
In the last line of Eq.~(\ref{num-coefficient}), we have used the resummed
result in Eq.~(\ref{resummed}). Taking $\alpha_s=\alpha_s(2m_c)=0.25$,
we see that the corrections of order $\alpha_s v^2$ and $\alpha_s v^4$ are
$0.5\%$ and $-0.2\%$, respectively. These are not very significant at the
current level of precision of the theory of $J/\psi$ decays to a lepton pair.

As can be seen from Eq.~(\ref{num-coefficient}), the velocity expansion
converges rapidly for approximate charmonium matrix elements.
In fact, the expressions for $\Delta G^{(1)}_{\overline{\rm MS}}$ in
Eq.~(\ref{DeltaGMS}) and $\Delta H^{(1)}$ in Eq.~(\ref{DeltaH}), taken as
functions of $v=|\bm{q}|/m$, have finite radii of convergence. The logarithms
in $L(\delta)$ [Eq.~(\ref{L-delta})] and the Spence functions in $K(\delta)$
[Eq.~(\ref{K-delta})] have branch points at $\delta=\pm 1$, i.e.,
$v=\pm\infty$. The quantity $\delta=v/\sqrt{1+v^2}$ has branch points at
$v=\pm i$. Therefore, the closest singularities to the origin in 
$\Delta G^{(1)}_{\overline{\rm MS}}$ or $\Delta H^{(1)}$ are at $v=\pm i$.
Consequently, the radii of convergence of $\Delta G^{(1)}_{\overline{\rm MS}}$
and  $\Delta H^{(1)}$ as functions of $v$ are one. It follows that the
velocity expansion for the $Q\bar Q$ operators is absolutely convergent,
provided that the absolute values of the operator matrix elements are bounded
by a geometric sequence in which the ratio between elements of the sequence
is less than $m^2$.
\section{Conclusions\label{sec:Conclusion}}
We have presented a calculation in NRQCD of the order-$\alpha_s$ corrections
to the quarkonium electromagnetic current. Our calculation gives expressions
for the short-distance coefficients of all of the $Q\bar Q$ NRQCD operators
that contain any number of derivatives but no gauge fields. 
Our operators are not gauge invariant, and we evaluate their matrix elements
in the Coulomb gauge. Our principal
results are given in Eqs.~(\ref{DeltaH}) and (\ref{DeltaGMS}). The NRQCD
short-distance coefficients can be obtained, according to
Eq.~(\ref{ab-MSbar}), from the Taylor-series expansions of
$\Delta G^{(1)}_{\overline{\rm MS}}$ in Eq.~(\ref{DeltaGMS}) and 
$\Delta H^{(1)}$ in Eq.~(\ref{DeltaH}). Our results at relative order $v^2$
agree with those in Ref.~\cite{Luke:1997ys}.

Our calculation makes use of a new method for computing, to all orders
in $v$, the one-loop NRQCD corrections that enter into the matching of
NRQCD to full QCD. In this new method, we begin with QCD expressions for
the loop integrands. We obtain the NRQCD corrections from these QCD
expressions by carrying out the integration over the temporal
component of the loop momentum and then  expanding the loop integrands
in powers of the loop and external momenta divided by the heavy-quark
mass $m$. We carry out this expansion {\it before} implementing the
dimensional regularization. The new approach allows one to avoid the
daunting task of obtaining NRQCD operators and interactions to all
orders in $v$, along with their Born-level short-distance coefficients,
and computing their contributions to the one-loop corrections. In terms
of the total labor involved, the computation of the NRQCD corrections
to all orders through the new approach is comparable to the
calculation of the NRQCD corrections at relative order $v^2$ through
conventional NRQCD methods. This new method should be applicable to
matching calculations for a variety of effective field theories,
including heavy-quark effective theory and soft-collinear effective
theory.

As we have mentioned, our approach is related to the method of regions
\cite{Beneke:1997zp}. The NRQCD corrections in our approach correspond
in the method of regions to the sum of the contributions from the
potential, soft, and ultrasoft regions, i.e., the contribution
from the small-loop-momentum region \cite{Beneke:1997zp}. In our
approach we have computed the quantities $\Delta G^{(1)}$ and $\Delta
H^{(1)}$ by subtracting the NRQCD corrections from the full-QCD
corrections. In the method of regions, $\Delta G^{(1)}$ and $\Delta
H^{(1)}$ could, in principle, be computed directly from the
contribution from the hard region. However, a straightforward
computation of the contribution from the hard region, carried out by
expanding the integrand in powers of the small momentum, would yield
Taylor-series expansions of $\Delta G^{(1)}$ and $\Delta H^{(1)}$ in
Eq.~(\ref{DeltaGH}) in powers of $\delta$. It would be nontrivial to
sum those expansions to obtain the compact expressions in
Eq.~(\ref{DeltaGH}). In contrast, in our approach, expansions of the
integrand occur only in the NRQCD expressions and lead to very simple
series that can be summed at the integrand level. Hence, our method may
be more efficient than the method of regions for computations of
short-distance coefficients to all orders in $v$. Our method is also 
applicable in the case of a hard-cutoff regulator, such as lattice 
regularization, while the method of regions applies only in the case of 
dimensional regularization.

Because we have omitted operators that contain gauge fields, the operators
that we consider are not the complete set of NRQCD operators that
describe the quarkonium electromagnetic current. In the Coulomb gauge,
the gauge-field
operators first enter at relative order $v^4$, and so our results cannot
be considered to be complete beyond order $v^2$. However, the operators
that we consider account for all of the contributions that are contained
in the Coulomb-gauge wave function of the quarkonium $Q\bar Q$ Fock state. 
The correction to the $S$-wave component of the electromagnetic current
that we find in relative order $\alpha_s v^4$ is
only about $-0.2$\%, which is not significant at the current level of 
the precision of the theory of $J/\psi$ decays to a lepton pair.

We have examined the convergence of the NRQCD velocity expansion for
$S$-wave $Q\bar{Q}$ operators. In Eq.~(\ref{num-coefficient}), we give the
numerical values for the sums of the first few $S$-wave contributions to
the electromagnetic current and for the sum of all of the $S$-wave
contributions. In these computations, we have made
use of the value of the relative-order-$v^2$ $J/\psi$
matrix element that is given in Ref.~\cite{Bodwin:2007fz} and the
approximate  relation
between operator matrix elements in Eq.~(\ref{gen-Gremm-Kapustin}),
which holds in spin-independent-potential models \cite{Bodwin:2006dn}.
It can be seen from Eq.~(\ref{num-coefficient}) that the velocity
expansion converges rapidly in this case. In fact, the expressions for 
$\Delta G^{(1)}_{\overline{\rm MS}}$ in Eq.~(\ref{DeltaGMS}) and
$\Delta H^{(1)}$ in Eq.~(\ref{DeltaH}), taken as functions of
$v=|\bm{q}|/m$, have radii of convergence one. Therefore, the velocity
expansion for the $Q\bar Q$ operators is absolutely convergent, provided
that the absolute values of the operator matrix elements are bounded by
a geometric sequence in which the ratio between elements of the sequence
is less than $m^2$.
\begin{acknowledgments}
The work of G.T.B. was supported by the U.S. Department of Energy, 
Division of High Energy Physics, under contract No.~DE-AC02-06CH11357. 
The work of H.S.C. was supported by the BK21 program. 
The work of C.Y. was supported by the Korea Research Foundation under 
MOEHRD Basic Research Promotion grant No.~KRF-2006-311-C00020.
The work of J.L. was supported by the Korea Science and Engineering
Foundation (KOSEF) funded by the Korea government (MEST) under 
grant No.~R01-2008-000-10378-0.
\end{acknowledgments}
\appendix

\section{Tensor-integral reduction\label{appendix:tensor}}
In this Appendix, we describe the tensor-integral reduction that we use
to simplify Eq.~(\ref{numerator}).

Tensor integrals of rank-1 and -2 that depend on $p$ or on
both $p$ and $q$ can be expressed in terms of scalar integrals as follows:
\begin{subequations}
\label{tensor-integral0}%
\begin{eqnarray}
\label{kmu-p-only}%
\int_k k^\mu f(k,p)&=&
\frac{p^\mu}{p^2}\int_k p\cdot k f(k,p),
\\
\int_k k^\mu k^\nu f(k,p)&=&
\int_k \left[
 d_1(k,p) g^{\mu\nu}
+d_2(k,p)p^\mu p^\nu \right] f(k,p),
\\
\int_k k^\mu f(k,p,q)&=&
p^\mu\int_k d_3(k,p,q) f(k,p,q)
+
q^\mu\int_k d_4(k,p,q) f(k,p,q),
\\
\int_k k^\mu k^\nu f(k,p,q)&=&
g^{\mu\nu}
\int_k d_5(k,p,q)  f(k,p,q)
+
p^\mu p^\nu 
\int_k d_6(k,p,q)  f(k,p,q)
\nonumber\\
&& +
q^\mu q^\nu
\int_k d_7(k,p,q)  f(k,p,q)
\nonumber\\
&& +
(p^\mu q^\nu + p^\nu q^\mu)
\int_k d_8(k,p,q)  f(k,p,q),
\end{eqnarray}
\end{subequations}
where $f$ is an arbitrary scalar function of the argument
four-vectors. If $p\cdot q=0$, then the functions $d_i$
in Eq.~(\ref{tensor-integral0}) are given by
\begin{subequations}
\label{tensor-integral-cd}%
\begin{eqnarray}
d_1(k,p)&=&\frac{1}{d-1}
\left[
k^2-\frac{(k\cdot p)^2}{p^2}
\right],
\\
d_2(k,p)&=&\frac{1}{(d-1)p^2}
\left[
-k^2+d\frac{(k\cdot p)^2}{p^2}
\right],
\\
d_3(k,p,q) 
&=& 
\frac{k\cdot p}{p^2},
\\
d_4(k,p,q) 
&=& 
\frac{k\cdot q}{q^2},
\\
d_5(k,p,q)
&=&
\frac{1}{d-2}
\left[
k^2
-\frac{(k\cdot p)^2}{p^2}
-\frac{(k\cdot q)^2}{q^2}
\right],
\\
d_6(k,p,q)
&=&
\frac{1}{(d-2)p^2}
\left[
-k^2
+(d-1)\frac{(k\cdot p)^2}{p^2}
+\frac{(k\cdot q)^2}{q^2}
\right],
\\
d_7(k,p,q)
&=&
\frac{1}{(d-2)q^2}
\left[
-k^2
+\frac{(k\cdot p)^2}{p^2}
+(d-1)\frac{(k\cdot q)^2}{q^2}
\right],
\\
d_8(k,p,q)
&=&
\frac{k\cdot p}{p^2}
\frac{k\cdot q}{q^2}.
\end{eqnarray}
\end{subequations}
\section{Integrals for the QCD corrections%
\label{appendix:scalarintegrals}}
In this Appendix, we evaluate the integrals in Eq.~(\ref{Si}).
Throughout this Appendix, we neglect expressions of order $\epsilon$
or higher.
The integrals in Eq.~(\ref{Si}) can be expressed in terms of elementary integrals
$I_{010}$, $I_{110}$, $I_{011}$, $I_{-111}$, and $I_{111}$:
\begin{subequations}
\label{JiI}%
\begin{eqnarray}
J_1&=&I_{011},
\\
J_2&=&I_{111},
\\
J_3&=&I_{110}-I_{011},
\label{JiI3}%
\\
J_4&=&\frac{1}{d-2}
\left( I_{011}-\frac{I_{-111}-I_{010}}{4q^2}
\right),
\\
J_5 &=& 0,
\\
J_6 &=& -J_4 + \frac{1}{4 q^2} I_{-1 1 1} 
+\frac{p^2 - 2 m^2}{4 q^2 m^2} I_{0 1 0},
\\
J_7 &=& 0,
\end{eqnarray}
\end{subequations}
where the scalar integral $I_{abc}$ is defined by
\begin{equation}
\label{Iabc}%
I_{abc}=
\int_k \frac{1}{D_0^a D_1^b D_2^c}.
\end{equation}
In deriving Eq.~(\ref{JiI}), we have used the fact that $I_{abc}=I_{acb}$,
which follows from the symmetry of the integrals under
$p_1\leftrightarrow p_2$ and $k\to -k$.
We have also discarded the scaleless, power-divergent integral $I_{100}$, which vanishes in
dimensional regularization.
In deriving the expressions for $J_4$ and $J_6$, we have made
a further tensor reduction, using Eq.~(\ref{kmu-p-only}),
which leads to
\begin{equation}
\label{I11m1}%
I_{1 1 -1}=
\frac{2p^2}{m^2}\,I_{010}.
\end{equation}

$I_{010}$ and $I_{110}$, which depend only on $m^2$, are given by
\begin{subequations}
\label{Ielementary}%
\begin{eqnarray}
I_{010}&=&
\frac{i}{(4\pi)^2}
\,\,m^2
\left(
\frac{1}{\epsilon_{\textrm{UV}}}
+\log\frac{4\pi\mu^2 e^{-\gamma_{_{\textrm{E}}}}}{m^2}
+1
\right),
\\
I_{110}&=&
\frac{i}{(4\pi)^2}
\left(
\frac{1}{\epsilon_{\textrm{UV}}}
+\log\frac{4\pi\mu^2 e^{-\gamma_{_{\textrm{E}}}}}{m^2}
+2
\right).
\end{eqnarray}
\label{simple}%
\end{subequations}

The scalar integrals $I_{0 1 1}$ and $I_{-1 1 1}$ 
can be evaluated by using
Feynman parametrization. After integrating over $k$, we obtain
\begin{subequations}
\label{IAB1}%
\begin{eqnarray}
I_{011}&=&
\frac{i}{(4\pi)^2}
\left(
\frac{4\pi \mu^2}{p^2}
\right)^\epsilon
\Gamma(\epsilon)
\int_0^1 dz \,( z^2-\delta^2-i\varepsilon )^{-\epsilon},
\\
I_{-111}&=&
\frac{i}{(4\pi)^2}
\left(
\frac{4\pi \mu^2}{p^2}
\right)^\epsilon
\Gamma(\epsilon)
\left(\frac{3-2\epsilon}{1-\epsilon}\right)
p^2 \int_0^1 dz \,( z^2-\delta^2-i\varepsilon )^{1-\epsilon},
\end{eqnarray}
\end{subequations}
where $z=2x-1$ and $x$ is the original Feynman parameter.
Expanding the integrands of Eq.~(\ref{IAB1}) in powers
of $\epsilon$, integrating over $z$, and using Eq.~(\ref{eq-delta}),
we find that
\begin{subequations}
\label{IABF}%
\begin{eqnarray}
I_{011}&=&
\frac{i}{(4\pi)^2}
\left[
\frac{1}{\epsilon_{\textrm{UV}}}
+\log\frac{4\pi\mu^2 e^{-\gamma_{_{\textrm{E}}}}}{m^2}
+2 -2\delta^2 L(\delta)
+i\pi\delta\, 
\right],
\\
I_{-111}&=&
\frac{i}{(4\pi)^2}
\,\,\frac{m^2}{1-\delta^2}
\bigg[
(1-3\delta^2)
\left(
\frac{1}{\epsilon_{\textrm{UV}}}
+\log\frac{4\pi\mu^2 e^{-\gamma_{_{\textrm{E}}}}}{m^2}
+1
\right)
-2\delta^2 
+4\delta^4 L(\delta)
-2\pi i\delta^3
\bigg],
\nonumber\\
\end{eqnarray}
\end{subequations}
where $L(\delta)$ is defined in Eq.~(\ref{L-delta}) and
we have used the following results, which hold for $0\le \delta<1$:
\begin{subequations}
\begin{eqnarray}
\int_0^1 \log(z^2-\delta^2-i\varepsilon)\,dz
&=&
-2 +2\delta^2 L(\delta)
+\log(1-\delta^2)
-i\pi\delta,\\
\int_0^1 (z^2-\delta^2-i\varepsilon)\log(z^2-\delta^2-i\varepsilon)\,dz
&=&
\frac{1}{3}
\bigg[
 -\frac{2}{3}
 +4\delta^2
 -4\delta^4 L(\delta)
\nonumber\\
&&
\hspace{3.1ex}
 +(1-3\delta^2)\log(1-\delta^2)
 +\,2\pi i\delta^3\,
 \bigg].
\end{eqnarray}
\end{subequations}

$I_{111}$ can be evaluated by using Feynman parametrization.
After integrating over $k$, we obtain
\begin{equation}
\label{IABC2}%
I_{111}=
- \frac{i}{(4\pi)^2}
\left(\frac{4\pi\mu^2}{p^2} \right)^\epsilon
\frac{\Gamma(1+\epsilon)}{p^2}
\int_0^1 dy \,y^{-1-2\epsilon}\int_0^{1}dz
\left(
z^2-\delta^2-i\varepsilon
\right)^{-1-\epsilon},
\end{equation}
where $z=2x-1$ and the original Feynman parameters are
$x$ and $y$. The infrared divergence is isolated in the integral over $y$:
\begin{equation}
\label{s-integral}%
\int_0^1 dy \, y^{-1-2\epsilon}=
-\frac{1}{2\epsilon_{\textrm{IR}}}.
\end{equation}
The integral over $z$ can be evaluated by expanding the integrand
in powers of $\epsilon$. Then, we obtain
\begin{eqnarray}
\label{IABCF}%
I_{111}&=&
\frac{i}{(4\pi)^2}
\,\, \frac{1-\delta^2}{4m^2}
\Bigg\{
        \bigg(
\frac{1}{\epsilon_{\textrm{IR}}}
+\log 
\frac{4 \pi \mu^2 e^{-\gamma_{_{\!\textrm{E}}}} }
     {m^2}
        \bigg)
\bigg[
-2L(\delta)
+\frac{i\pi}{\delta}
\bigg]
+4K(\delta)
\nonumber\\
&&
\hspace{17ex}
-\frac{\pi^2}{\delta}
-\frac{i\pi}{\delta}
 \log\frac{4\delta^2}{1-\delta^2}
\Bigg\},
\end{eqnarray}
where $K(\delta)$ is defined in Eq.~(\ref{K-delta}) and
we have used the following results, which hold for $0\le \delta<1$:
\begin{subequations}
\begin{eqnarray}
\int_0^{1}
\frac{dz}{
z^2-\delta^2-i\varepsilon}
&=&
\frac{i\pi}{2\delta}-L(\delta),
\\
\int_0^{1}
\frac{\log(z^2-\delta^2-i\varepsilon)}{
z^2-\delta^2-i\varepsilon}\,dz
&=&
-
\log(1-\delta^2)L(\delta)
-2 K(\delta)
+\frac{\pi^2}{2\delta}
+\frac{i\pi}{\delta} \log(2\delta).
\end{eqnarray}
\end{subequations}

By making use of Eqs.~(\ref{JiI}), (\ref{Ielementary}),
(\ref{IABF}), and (\ref{IABCF}), we find that
\begin{subequations}
\label{Ji-final}%
\begin{eqnarray}
J_1&=&
\frac{i}{(4\pi)^2}
\bigg[
\frac{1}{\epsilon_{\textrm{UV}}}
+\log 
\frac{4 \pi \mu^2 e^{-\gamma_{_{\!\textrm{E}}}}}{m^2}
+2
-2\delta^2 L(\delta)
+i\pi\delta
\bigg],
\\
J_2&=&
\frac{i}{(4\pi)^2}
\frac{1-\delta^2}{4m^2}
\Bigg\{
        \bigg(
\frac{1}{\epsilon_{\textrm{IR}}}
+\log
\frac{4 \pi \mu^2 e^{-\gamma_{_{\!\textrm{E}}}} }
     {m^2}
        \bigg)
\bigg[
-2L(\delta)
+\frac{i\pi}{\delta}
\bigg]
+4K(\delta)
\nonumber\\
&&
\hspace{17ex}
-\frac{\pi^2}{\delta}
-\frac{i\pi}{\delta}
 \log\frac{4\delta^2}{1-\delta^2}
\Bigg\},
\\
J_3&=&
\frac{i}{(4\pi)^2}
\bigg[
2\delta^2 L(\delta)
-i\pi\delta
\bigg],
\\
J_4
&=&
\frac{1}{4}
\left[
\frac{i}{(4\pi)^2}
+J_1\right],
\\
J_5 &=& 0,
\\
J_6 &=&
-\frac{1}{4}J_3,
\\
J_7 &=& 0.
\end{eqnarray}
\end{subequations}
The results for $J_1$--$J_4$ in Eq.~(\ref{Ji-final}) agree with those in
Ref.~\cite{Lee:2007kg}.
\section{Integrals for the NRQCD corrections\label{appendix:IntS}}
Here, we tabulate some integrals that are useful in computing the
NRQCD corrections.

In dimensional regularization, scaleless, power-divergent
integrals vanish:
\begin{equation}
\label{scaleless0}%
\int_{\bm{k}} \frac{1}{|\bm{k}|^{n}}
=0
\end{equation}
for $n\neq 3$.
The only scaleless logarithmically divergent integral that we
encounter is
\begin{equation}
\label{n0}%
n_0\equiv
\int_{\bm{k}} \frac{1}{|\bm{k}|^3}
=\frac{1}{4\pi^2}\left(
 \frac{1}{\epsilon_{\textrm{UV}}}
-\frac{1}{\epsilon_{\textrm{IR}}}\right).
\end{equation}
There are a few integrals that depend on $\bm{q}$ that appear
in the evaluations of the $S_i$ in Eq.~(\ref{Si-definition}):
\begin{subequations}
\begin{eqnarray}
\label{n1}%
n_1&\equiv&
\int_{\bm{k}}
\frac{1}{\bm{k}^2+2\bm{k}\cdot \bm{q}-i\varepsilon}
=\frac{i}{4\pi}|\bm{q}|,
\\
\label{n2}%
n_2&\equiv&
\int_{\bm{k}} 
\frac{1}{\bm{k}^2(\bm{k}^2+2\bm{k}\cdot \bm{q}-i\varepsilon)}
\nonumber\\
&=&-\frac{i}{16\pi|\bm{q}|} \left( 
\frac{1}{\epsilon_{\textrm{IR}}}
+\log 
\frac{\pi \mu^2 e^{-\gamma_{_{\!\textrm{E}}}} }{\bm{q}^2}
+i\pi
                  \right),
\\
\label{n3}%
n_3&\equiv&
\int_{\bm{k}}
\frac{\bm{k}^2}{\bm{k}^2+2\bm{k}\cdot \bm{q}-i\varepsilon}
=\frac{i}{2\pi}|\bm{q}|^3.
\end{eqnarray}
\end{subequations}

We also make use of the angular averages:
\begin{subequations}
\label{angularav}%
\begin{eqnarray}
\label{angularav1}%
\int_{\bm{k}}
\frac{f(\bm{k}^2)}{E\pm\bm{q}\cdot\hat{\bm{k}}}
&=&
\frac{1}{2|\bm{q}|} \log \left(\frac{E+|\bm{q}|}{E-|\bm{q}|} \right)
\int_{\bm{k}}f(\bm{k}^2),
\\
\label{angularav2}%
\int_{\bm{k}} 
\frac{f(\bm{k}^2)}{E^2 - (\bm{q}\cdot\hat{\bm{k}})^2}
&=&
\frac{1}{2E}
\int_{\bm{k}} f(\bm{k}^2) \left( 
\frac{1}{E+ \bm{q}\cdot\hat{\bm{k}}}
+
\frac{1}{E- \bm{q}\cdot\hat{\bm{k}}}
\right)
\nonumber\\
&=&
\frac{1}{2E|\bm{q}|} \log \left(\frac{E+|\bm{q}|}{E-|\bm{q}|} \right)
\int_{\bm{k}}f(\bm{k}^2),
\end{eqnarray}
\end{subequations}
where $f(\bm{k}^2)$  is any function of $\bm{k}^2$.

\section{Evaluation of the integrals for
$\bm{[Z_Q]}_{\textbf{NRQCD}}$\label{appendix:T-integrals}}
In this Appendix, we evaluate 
the integrals $T_{02}$, $T_{11}$, $T_{12}$, $T_{02}^\mu$, $T_{11}^\mu$,
and $T_{12}^{\mu}$,
which enter into the calculation of 
$[Z_Q]_{\textrm{NRQCD}}$ and are defined in Eq.~(\ref{ti-definition}).
We make use of the same strategy that we used in evaluating the
$S_i$ integrals in Sec.~\ref{sec:NRQCD}, except that we 
carry out the evaluation in
the rest frame of the heavy quark, $p_1=(m,\bm{0})$, where the 
expressions become compact.
The change of frame shifts momenta by an amount of order $mv$.  
Therefore, the NRQCD expansion in powers of the external momentum
divided by $m$ remains valid. In the heavy-quark rest frame, the gluon-
and quark-propagator denominators are
\begin{subequations}
\label{dp-rest}%
\begin{eqnarray}
{[D_0]}_{\textrm{ rest}}&=&
(k^0+|\bm{k}|-i\varepsilon)
(k^0-|\bm{k}|+i\varepsilon),
\\
{[D_1]}_{\textrm{ rest}}&=&
(k^0+m+\sqrt{m^2+\bm{k}^2}-i\varepsilon)
(k^0+m-\sqrt{m^2+\bm{k}^2}+i\varepsilon).
\end{eqnarray}
\end{subequations}
We evaluate the $k^0$ integrals by using contour integration,
closing the contour in the upper half-plane in every case.
We denote the contributions of gluon and quark poles
by subscripts $g$ and $Q$, respectively.

The integral $T_{02Q}$ is
\begin{equation}
\label{t0q}%
T_{02Q}
=
\frac{i}{4}\intN\frac{1}
{(m^2+\bm{k}^2)^{3/2}}.
\end{equation}

The integral $T_{11}$ yields
\begin{subequations}
\label{t11k}%
\begin{eqnarray}
T_{11g}
&=&
\frac{i}{4m}\int_{\bm{k}}\frac{1}{\bm{k}^2},
\\
T_{11Q}
&=&
-\frac{i}{4m}\intN\frac{1}{\bm{k}^2}
\left(
1-\frac{m}{\sqrt{m^2+\bm{k}^2}}
\right).
\end{eqnarray}
\end{subequations}

The integral $T_{12}$ yields
\begin{subequations}
\label{t12k}%
\begin{eqnarray}
\label{t12g}%
T_{12g}
&=&
-\frac{i}{8m^2}\int_{\bm{k}}\frac{1}{|\bm{k}|^3}=
-\frac{i}{8m^2}\,n_0,
\\
T_{12Q}
&=&
\frac{i}{8m^2}\intN\frac{1}
{(m^2+\bm{k}^2)^{3/2}},
\end{eqnarray}
\end{subequations}
where $n_0$ is defined in Eq.~(\ref{n0}).

In the cases of the integrals $T^\mu_{02}$, $T^{\mu}_{11}$, and $T^{\mu}_{12}$,
the integrand of the temporal component $T^0_{ab}$ is identical to that of 
$T_{ab}$, except that the integrand in $T^0_{ab}$ contains an additional factor 
of $k^0$. Integrating over $k^0$, we obtain 
\begin{subequations}
\label{t0ab}%
\begin{eqnarray}
T_{02Q}^0 &=&
 -\frac{i m}{4} \intN \frac{1}{(m^2+\bm{k}^2)^{3/2}},
\\
T_{11 g}^0 &=& -\frac{i}{4 m} \int_{\bm{k}} \frac{1}{|\bm{k}|},
\\
T_{11 Q}^0 &=& \frac{i}{4 m} \intN \frac{1}{\sqrt{m^2+\bm{k}^2}},
\\
T^{0}_{12g}
&=&
\frac{i}{8m^2}\int_{\bm{k}}\frac{1}{|\bm{k}|^2},
\\
T^{0}_{12Q}
&=&
-\frac{i}{8m^2}\intN\frac{1}{\bm{k}^2}
\left[1-\frac{m^{3}}{(m^2+\bm{k}^2)^{3/2}}\right].
\end{eqnarray}
\end{subequations}
For the spatial component ${T}^i_{ab}$, the integrand is identical 
to the integrand in $T_{ab}$, except that the integrand in ${T}^i_{ab}$ 
contains an additional factor ${k}^i$.  
By making use of Eqs.~(\ref{t0q})--(\ref{t12k}), 
we find that
\begin{subequations}
\label{tiab}%
\begin{eqnarray}
{T}^i_{02Q} 
&=& \frac{i}{4} \intN \frac{{k}^i}{(m^2+\bm{k}^2)^{3/2}},
\\
{T}^i_{11 g} &=& \frac{i}{4 m} \int_{\bm{k}} \frac{{k}^i}{\bm{k}^2},
\\
{T}^i_{11 Q} &=&-\frac{i}{4 m} \intN
\frac{{k}^i}{\bm{k}^2}
\left(
1-\frac{m}{\sqrt{m^2+\bm{k}^2}}
\right),
\\
{T}^i_{12g}
&=&
-\frac{i}{8m^2}\int_{\bm{k}}\frac{{k}^i}{|\bm{k}|^3},
\\
{T}^i_{12Q}
&=&
\frac{i}{8m^2}\intN\frac{{k}^i}
{(m^2+\bm{k}^2)^{3/2}}.
\end{eqnarray}
\end{subequations}

Expanding the integrands in Eqs.~(\ref{t0q})--(\ref{tiab}) 
in powers of $\bm{k}^2/m^2$, we find that
all of the terms in the expansions yield scaleless, power-divergent
integrals, with the exception of the integral $T_{12g}$ in 
Eq.~(\ref{t12g}).
Therefore,
\begin{equation}
\label{t-vanishing}%
T_{11}=T_{02}=T^{\mu}_{02}=T^\mu_{11} = T^\mu_{12}=0
\end{equation}
and
\begin{equation}
\label{t12f}%
T_{12}=-\frac{i}{32\pi^2 m^2}
  \left(
 \frac{1}{\epsilon_{\textrm{UV}}}
-\frac{1}{\epsilon_{\textrm{IR}}}\right),
\end{equation}
where we have used Eq.~(\ref{n0}).


\begin{thebibliography}{}
\bibitem{Bodwin:2007fz}
  G.~T.~Bodwin, H.~S.~Chung, D.~Kang, J.~Lee, and C.~Yu,
  Phys.\ Rev.\  D {\bf 77}, 094017 (2008)
  [arXiv:0710.0994 [hep-ph]].
\bibitem{Van Royen:1967nq}
  R.~Van Royen and V.~F.~Weisskopf,
  Nuovo Cim.\  A {\bf 50}, 617 (1967)
  [Erratum-ibid.\  A {\bf 51}, 583 (1967)].
\bibitem{Barbieri:1975ki}
  R.~Barbieri, R.~Gatto, R.~Kogerler, and Z.~Kunszt,
  Phys.\ Lett.\ B {\bf 57}, 455 (1975).
\bibitem{Celmaster:1978yz}
  W.~Celmaster,
  Phys.\ Rev.\  D {\bf 19}, 1517 (1979).
\bibitem{Bodwin:1994jh}
  G.~T.~Bodwin, E.~Braaten, and G.~P.~Lepage,
  Phys.\ Rev.\  D {\bf 51}, 1125 (1995)
  [Erratum-ibid.\  D {\bf 55}, 5853 (1997)]
  [arXiv:hep-ph/9407339].
\bibitem{Bodwin:2002hg}
  G.~T.~Bodwin and A.~Petrelli,
  Phys.\ Rev.\  D {\bf 66}, 094011 (2002)
  [arXiv:hep-ph/0205210].
\bibitem{Czarnecki:1997vz}
  A.~Czarnecki and K.~Melnikov,
  Phys.\ Rev.\ Lett.\  {\bf 80}, 2531 (1998)
  [arXiv:hep-ph/9712222].
\bibitem{Beneke:1997jm}
  M.~Beneke, A.~Signer, and V.~A.~Smirnov,
  Phys.\ Rev.\ Lett.\  {\bf 80}, 2535 (1998)
  [arXiv:hep-ph/9712302].
\bibitem{Luke:1997ys}
  M.~E.~Luke and M.~J.~Savage,
  Phys.\ Rev.\  D {\bf 57}, 413 (1998)
  [arXiv:hep-ph/9707313].
\bibitem{Beneke:1997zp}
  M.~Beneke and V.~A.~Smirnov,
  Nucl.\ Phys.\  B {\bf 522}, 321 (1998)
  [arXiv:hep-ph/9711391].
\bibitem{Bodwin:1998mn}
  G.~T.~Bodwin and Y.~Q.~Chen,
  Phys.\ Rev.\  D {\bf 60}, 054008 (1999)
  [arXiv:hep-ph/9807492].
\bibitem{Braaten:1995ej}
  E.~Braaten and S.~Fleming,
  Phys.\ Rev.\  D {\bf 52}, 181 (1995)
  [arXiv:hep-ph/9501296].
\bibitem{Brambilla:2006ph}
  N.~Brambilla, E.~Mereghetti, and A.~Vairo,
  JHEP {\bf 0608}, 039 (2006)
  [arXiv:hep-ph/0604190].
\bibitem{Bodwin:2006dn}
  G.~T.~Bodwin, D.~Kang, and J.~Lee,
  Phys.\ Rev.\ D {\bf 74}, 014014 (2006)
  [arXiv:hep-ph/0603186].
\bibitem{Gremm:1997dq}
  M.~Gremm and A.~Kapustin,
  Phys.\ Lett.\  B {\bf 407}, 323 (1997)
  [arXiv:hep-ph/9701353].
\bibitem{Lee:2007kg}
  J.~Lee, H.~K.~Noh, and C.~Yu,
  J.\ Korean Phys.\ Soc.\  {\bf 50}, 403 (2007).
\end{thebibliography}
\end{document}